\documentclass[11pt]{article}

\usepackage[final]{acl}

\usepackage{times}
\usepackage{latexsym}

\usepackage[T1]{fontenc}

\usepackage[utf8]{inputenc}

\usepackage{microtype}

\usepackage{inconsolata}

\usepackage{graphicx}

\usepackage{amssymb}
\usepackage{amsmath}
\usepackage{algorithm}
\usepackage{algorithmic}
\usepackage{multirow}
\usepackage{tabularx, booktabs}
\usepackage{xcolor}
\usepackage[table]{xcolor}
\usepackage{subcaption}
\newcolumntype{Y}{>{\centering\arraybackslash}X}
\newcolumntype{L}{>{\raggedright\arraybackslash}X}

\newtheorem{theorem}{Theorem}[section]

\usepackage{enumitem}
\usepackage{siunitx}
\usepackage[table]{xcolor}
\usepackage{appendix}
\usepackage{bm}
\definecolor{lightblue}{HTML}{D7F6FF} 
\definecolor{midgray}{HTML}{E6E6E6}  
\definecolor{orange}{HTML}{EC805A}
\definecolor{sensitiveColor}{RGB}{200, 0, 0}   
\definecolor{meaningfulColor}{RGB}{110, 110, 110} 
\definecolor{SpanYellow}{HTML}{FFFF00} 
\definecolor{SpanBlue}{HTML}{0D20D0}   
\definecolor{SpanRed}{HTML}{C00000}    

\DeclareRobustCommand{\hlcap}[2]{%
  \begingroup
  \setlength{\fboxsep}{0.25ex}%
  \colorbox{#1}{\strut #2}%
  \endgroup
}

\usepackage{tikz}
\usepackage[utf8]{inputenc}

\newcommand{\blackcircnum}[1]{%
  \tikz[baseline=-0.75ex]{
    \node[
      shape=circle,
      draw=black,
      fill=black,
      text=white,
      inner sep=0.5pt,
      font=\scriptsize\bfseries,
      minimum size=0.8em
    ] (char) {#1};
  }%
}

\newcommand{\sen}[1]{\textcolor{sensitiveColor}{\textit{#1}}} 
\newcommand{\neu}[1]{\textcolor{meaningfulColor}{\textit{#1}}} 
\newcommand{\ctx}[1]{\textit{#1}} 

\usepackage{graphicx}
\usepackage{adjustbox}
\usepackage{subcaption}
\usepackage[most]{tcolorbox}
\usepackage{xcolor}

\title{Mask-Free Privacy Extraction and Rewriting: A Domain-Aware Approach via Prototype Learning}

\author{
\textbf{Xiaodong Li}$^{1,2*}$, \textbf{Yuhua Wang}$^{2,3*}$, \textbf{Qingchen Yu}$^{2,3}$, \textbf{Zixuan Qin}$^{1}$,\\
\textbf{Yifan Sun}$^{1,2\dagger}$, \textbf{Qinnan Zhang}$^{2,3\dagger}$, \textbf{Hainan Zhang}$^{2,3}$, \textbf{Zhiming Zheng}$^{2,3}$\\[4pt]
$^{1}$Center for the Applied Statistics, School of Statistics, Renmin University of China \\
$^{2}$Beijing Advanced Innovation Center for Future Blockchain and Privacy Computing\\
$^{3}$School of Artificial Intelligence, Beihang University, China\\[4pt]
\texttt{muzi\_4@ruc.edu.cn, yuhuawang@buaa.edu.cn}
}

\begin{document}
\maketitle 
\begin{abstract} 
Client-side privacy rewriting is crucial for deploying LLMs in privacy-sensitive domains. However, existing approaches struggle to balance privacy and utility. 
Full-text methods often distort context, while span-level approaches rely on impractical manual masks or brittle static dictionaries. 
Attempts to automate localization via prompt-based LLMs prove unreliable, as they suffer from unstable instruction following that leads to privacy leakage and excessive context scrubbing.
To address these limitations, we propose \textbf{DAMPER} (\textbf{D}omain-\textbf{A}ware \textbf{M}ask-free \textbf{P}rivacy \textbf{E}xtraction and \textbf{R}ewriting). DAMPER operationalizes latent privacy semantics into compact \textbf{Domain Privacy Prototypes} via contrastive learning, enabling precise, autonomous span localization. Furthermore, we introduce a Prototype-Guided Preference Alignment, which leverages learned prototypes as semantic anchors to construct preference pairs, optimizing a domain-compliant rewriting policy without human annotations. At inference time, DAMPER integrates a sampling-based Exponential Mechanism to provide rigorous span-level Differential Privacy (DP) guarantees. Extensive experiments demonstrate that DAMPER significantly outperforms existing baselines, achieving a superior privacy-utility trade-off.
\end{abstract}

\begingroup
\renewcommand{\thefootnote}{\fnsymbol{footnote}}
\footnotetext[1]{Equal contribution.}
\footnotetext[2]{Corresponding author.}
\endgroup


\section{Introduction}
\vspace{-1mm}
Deploying cloud-hosted Large Language Models (LLMs) in privacy-sensitive domains (e.g., healthcare, law, finance) often necessitates transmitting sensitive queries~\cite{qin2025achillesheelllmsaltering, yu-etal-2025-guessarena}, posing privacy risks under untrusted providers~\cite{staab2024beyond, kim2023propile}.
\emph{Client-side privacy rewriting} mitigates this by locally sanitizing content while preserving task intent~\cite{shi-etal-2022-just,igamberdiev-etal-2022-dp}. 
 
\begin{figure}[t]
\centering  \includegraphics[width=1\linewidth]{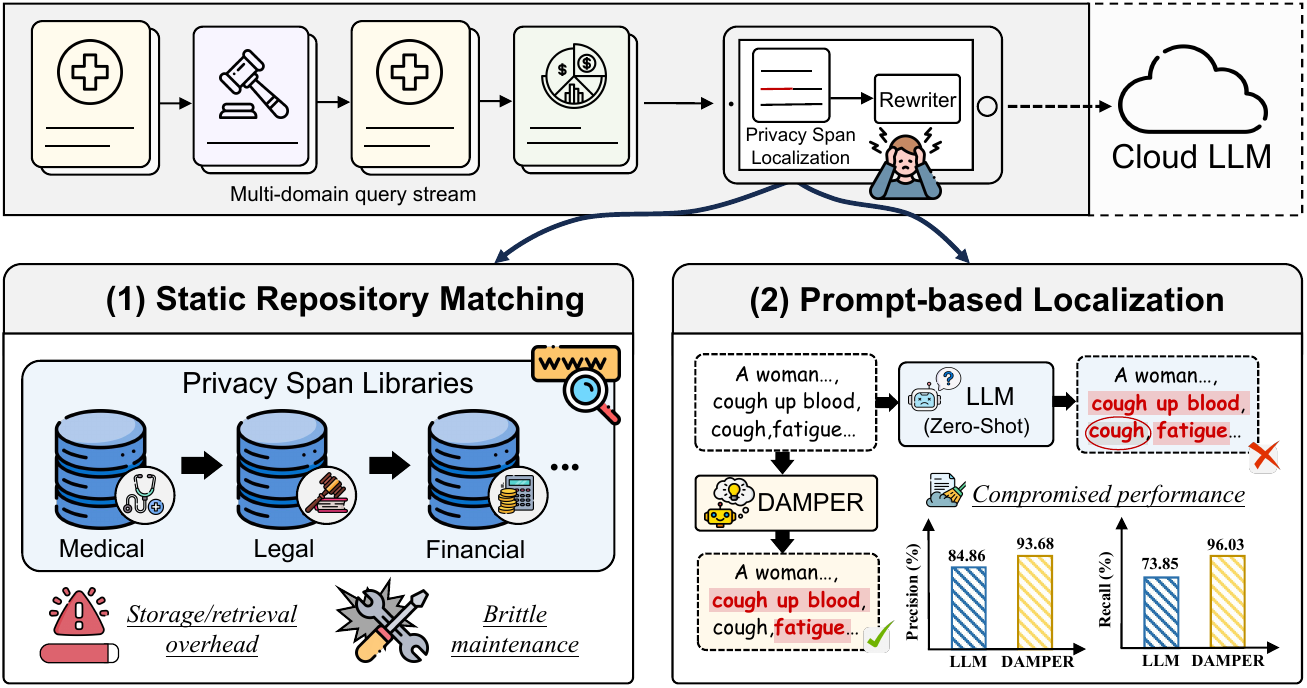} 
\caption{
\textbf{Challenge illustration.} (1) Static repository matching requires distinct privacy libraries for each domain, incurring excessive storage and maintenance overheads. (2) Prompt-based localizations lack granularity, resulting in compromised performance: they indiscriminately mask generic terms (low Precision) while missing actual sensitive ones (suboptimal Recall). 
Conversely, DAMPER leverages domain prototypes to discern context, achieving superior performance.
}
\label{fig:motivation}
\end{figure}

Early approaches~\cite{mattern-etal-2022-limits,utpala-etal-2023-locally,meisenbacher-etal-2024-dp} primarily operated at the \emph{full-text level}, rewriting the entire input and uniformly applying Differential Privacy (DP) mechanisms~\citep{dwork2006calibrating}. While straightforward, such coarse-grained perturbations often distort non-sensitive context, eroding domain style and professionalism and degrading downstream utility. 
This has motivated \emph{span-level privacy rewriting}~\cite{huang2025zeroshot,zeng-etal-2025-privacyrestore}, which selectively rewrites only privacy spans—contiguous segments that encode sensitive attributes—while preserving the remaining context verbatim. 
However, many span-level methods \textbf{implicitly assume that users can provide privacy masks or otherwise specify which spans are sensitive}. 
This assumption rarely holds in practical client deployment, where query streams span multiple professional domains and privacy semantics are inherently domain-contingent.
Although users typically know what they want to ask, they often lack the expertise to localize privacy spans according to domain-specific standards. 
Therefore, a practical rewriter must autonomously infer domain context and localize privacy spans, a capability we term \textit{mask-free} rewriting.

Two natural strategies for mask-free localization fall short in practice.
\textbf{\blackcircnum{1} Static Repository Matching.} Supporting multi-domain streams requires maintaining separate privacy span libraries, incurring significant storage and retrieval overheads. Moreover, the open-ended nature of sensitive vocabulary makes such maintenance brittle and costly.
\textbf{\blackcircnum{2} Prompt-based Localization.} Even when provided with detailed prompts strictly aligned with the privacy definitions used for dataset annotation (see Appendix~\ref{app:template_prompt}), zero-shot local LLMs struggle with span-level instruction following. As detailed in our case studies in Appendix~\ref{app:QualitativeAnalysis}, this instability leads to simultaneous false positives and negatives. Such errors either \textit{over-scrub} benign context, thereby degrading downstream utility, or leave privacy spans \textit{exposed}, directly compromising user privacy. Ultimately, this corrupts the evidentiary basis required for reliable downstream reasoning.
 
These limitations call for a lightweight, domain-aware framework capable of capturing latent privacy concepts to enforce domain-consistent boundaries.
This setting, where both domain semantics and privacy boundaries are latent at inference time, raises two fundamental questions:
\textbf{Q1.} \textit{How can we operationalize domain-specific privacy semantics to enable autonomous span localization without manual masks?}
\textbf{Q2.} \textit{How can we learn a utility-preserving rewriting policy without human annotations, while strictly satisfying DP constraints?}

In this work, we propose \textbf{D}omain-\textbf{A}ware \textbf{M}ask-Free \textbf{P}rivacy \textbf{E}xtraction and \textbf{R}ewriting, abbreviated as \textbf{DAMPER}. DAMPER operationalizes domain privacy semantics with \textbf{Domain Privacy Prototypes}—compact latent representations that guide the full lifecycle from localization to rewriting. Our key intuition is that \emph{\textbf{domain matters twice}}: it determines \emph{\textbf{which}} spans are privacy-sensitive, and it constrains \emph{\textbf{how}} they should be rewritten to maintain professional consistency.
To address \textbf{Q1}, we introduce a prototype-driven localization mechanism. We learn discriminative span representations using a multi-domain contrastive objective that pulls together private spans within the same domain while pushing apart non-private text and private spans from other domains. We then cluster these embeddings to obtain Domain Privacy Prototypes, which compress diverse and sensitive instances into a small set of semantic anchors. At inference time, DAMPER computes affinities between input segments and the learned prototypes to infer the dominant domain and localize privacy spans \emph{without} manual masks.
To tackle \textbf{Q2}, we propose a preference-based rewriting paradigm that explicitly manages the tension between obfuscation and context preservation.
Rather than imitating static references, we employ Prototype-Guided Direct Preference Optimization (DPO)~\citep{ouyang2022training} with automatically constructed preference pairs that penalize rewrites overly similar to the original sensitive span (privacy) while rewarding consistency with the domain prototype (utility).
This yields a domain-compliant rewriting policy \emph{without} human preference annotations.
During inference, we integrate a sampling-based Exponential Mechanism into token generation to provide span-level DP guarantees. 
Our contributions are:
\begin{itemize}[itemsep=1pt, topsep=1pt, leftmargin=10pt]
\item We propose a domain-aware rewriting framework that decouples sanitization from downstream LLMs, facilitating autonomous, mask-free span localization across multi-domains.
\item We introduce a prototype-guided preference alignment paradigm that leverages learned prototypes as consistency anchors, explicitly navigating the conflict between obfuscation and preservation in an annotation-free manner. 
\item We derive a span-level DP guarantee via the Exponential Mechanism, and demonstrate through extensive experiments that DAMPER achieves superior privacy--utility trade-offs. 
\end{itemize}

\vspace{-1mm}
\section{Related Work}
\vspace{-1mm}
\textbf{Token-level LDP Substitution.}
Early approaches~\citep{yue-etal-2021-differential, chen-etal-2023-customized} operate under metric local differential privacy, replacing each word with a neighbor in the embedding space.
While providing formal guarantees, these methods treat every token as equally sensitive. This indiscriminate perturbation often destroys syntactic structure and semantic utility, as noise is injected regardless of the token's privacy contribution.

\vspace{0.5mm}
\noindent\textbf{Sequence-level DP Paraphrasing.} To improve coherence, sequence-level methods~\citep{mattern-etal-2022-limits,igamberdiev-habernal-2023-dp,utpala-etal-2023-locally,meisenbacher-etal-2024-dp} rewrite entire texts using encoder-decoder architectures or LLM prompting. Although they generate more fluent outputs than token substitution, they apply a uniform privacy budget across the input. This lack of granularity forces a trade-off: sufficient noise to protect sensitive details often obfuscates the general context necessary for downstream tasks.

\noindent\textbf{Span-level Privacy Rewriting.} Recent frameworks adopt a detect-then-sanitize paradigm, utilizing strategies such as server-side restoration~\citep{zeng-etal-2025-privacyrestore}, zero-shot rewriting~\citep{huang2025zeroshot}, or inference-time distribution mixing~\citep{thareja2025dp}. However, these methods predominantly rely on rigid detection mechanisms (e.g., static inventories or explicit user masks), lacking the autonomous identification capability essential for open-ended domain semantics. Structurally, they often face a dilemma: search-based approaches typically lack formal DP guarantees, while restoration-dependent methods are tightly coupled to specific server-side LLMs, limiting model-agnostic deployment. 

Full discussions of related work and preliminaries are deferred to Appendix~\ref{app:related_work} and~\ref{sec:prelim}, respectively.

\vspace{-1mm}
\section{Problem Formulation}
\label{sec:problem}
\vspace{-1mm}
\noindent\textbf{Threat Model.}
We consider a client-side setting where the user’s device runs a local privacy rewriter before sending a query to an untrusted cloud LLM. The adversary is the cloud provider, or any party observing the transmitted query, with full access to the rewritten text $y$ and arbitrary inference or auxiliary knowledge to recover sensitive content from the original input $x$. In particular, the adversary may try to reconstruct privacy spans or infer private attributes implied by them.

\vspace{0.5mm}
\noindent\textbf{Span Definitions.}
We view a user query $x$ as a sequence of tokens containing a set of semantically meaningful spans, denoted as $\mathcal{U}$. Within this universal set, we distinguish a subset of privacy spans $\mathcal{S} \subset \mathcal{U}$ that convey sensitive information (e.g., detailed clinical symptoms). The remaining segments in $\mathcal{U}$ are considered non-privacy spans—meaningful but explicitly non-sensitive phrases. Any tokens not belonging to $\mathcal{U}$ (e.g., functional connectives or generic terms) are treated as ordinary context. 
For example, in the medical text ``\ctx{A woman \dots cough, cough up blood, fatigue \dots}'', the set $\mathcal{U}$ includes meaningful segments like \ctx{cough}, \ctx{cough up blood}, and \ctx{fatigue}. Here, $\mathcal{S}$ consists of \sen{cough up blood} and \sen{fatigue}, while \neu{cough} remains in $\mathcal{U}$ as a non-privacy span. Generic phrases like \ctx{A woman} fall outside $\mathcal{U}$ as ordinary context.

\vspace{0.5mm}
\noindent\textbf{Task Definition.} 
Given an input $x$, we aim to learn a rewriter $\pi_\theta$ that maps $x$ to a rewritten query $y$. In open-world settings where both the target domain and privacy annotations $\mathcal{S}$ are latent, the proposed method must satisfy three objectives:
1) \textbf{Domain-Aware Localization}. Dynamically infer the domain context of $x$ to localize the privacy spans $\hat{\mathcal{S}} \approx \mathcal{S}$ within $\mathcal{U}$, distinct from non-privacy spans.
2) \textbf{DP Rewriting}. Generate rewritten replacements for the detected spans $\hat{\mathcal{S}}$ under formal DP guarantees to obfuscate sensitive attributes.
3) \textbf{Utility Preservation}. Preserve the semantic integrity of the non-sensitive context ($\mathcal{U} \setminus \hat{\mathcal{S}}$) and the stylistic consistency of the domain, ensuring the rewritten query remains effective for downstream tasks.

\vspace{-1mm}
\section{Methodology} 
\begin{figure*}[t]
\centering
\includegraphics[width=\textwidth]{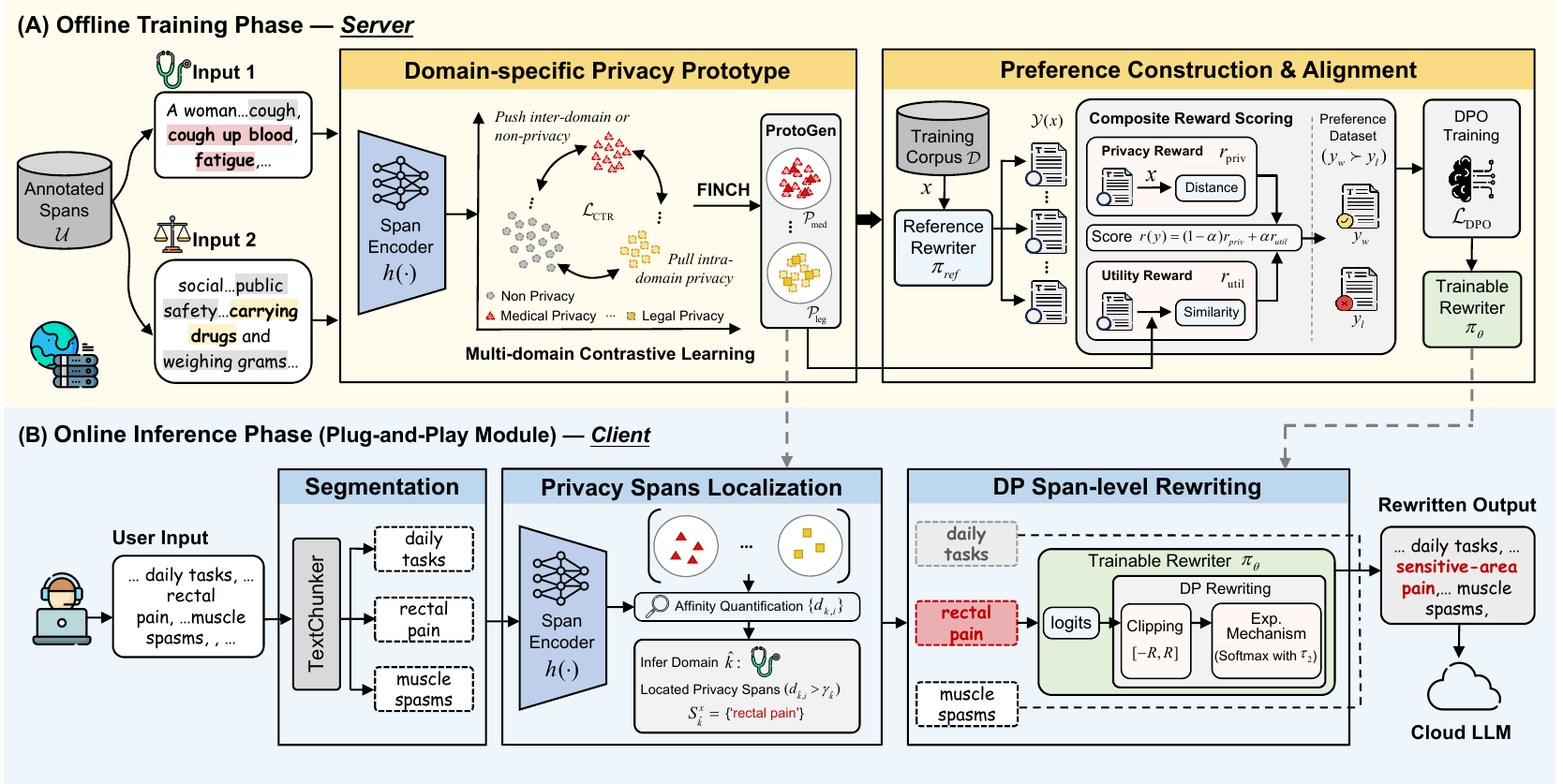}
\caption{\textbf{Overview of the DAMPER}. (A) Offline Training Phase (Sec.~\ref{sec:offline}):  We first employ multi-domain contrastive learning to cluster annotated spans into compact \textit{Domain Privacy Prototypes} (e.g., $\mathcal{P}_{med}$, $\mathcal{P}_{leg}$). These prototypes guide DPO training for rewriter $\pi_\theta$ using a composite reward to balance semantic obfuscation and domain fidelity. (B) Online Inference Phase (Sec.~\ref{sec:online}): The module segments user input via TextChunker and dynamically infers the target domain to localize privacy spans based on prototype affinity. Detected spans are rewritten using a sampling exponential mechanism, preserving non-sensitive context verbatim.}
\label{fig:overview}
\end{figure*}

\vspace{-1mm}
As shown in Figure~\ref{fig:overview}, DAMPER operates in two phases. The Offline Training constructs Domain Privacy Prototypes to encode domain privacy and align the rewriting policy via preference learning. The Online Inference serves as a plug-and-play module for domain-adaptive span localization and differentially private rewriting of sensitive regions. Appendix~\ref{app:algorithm} presents the algorithms of DAMPER.

\vspace{-1mm}
\subsection{Offline Training Phase} 
\label{sec:offline}
\vspace{-1mm}

\begin{figure}[t]
    \centering
    \begin{adjustbox}{width=0.45\textwidth,center}  
        \includegraphics{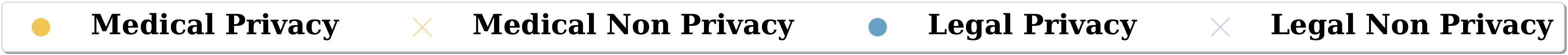}  
    \end{adjustbox}
    \begin{subfigure}[b]{0.20\textwidth}
        \centering
        \includegraphics[width=\textwidth]{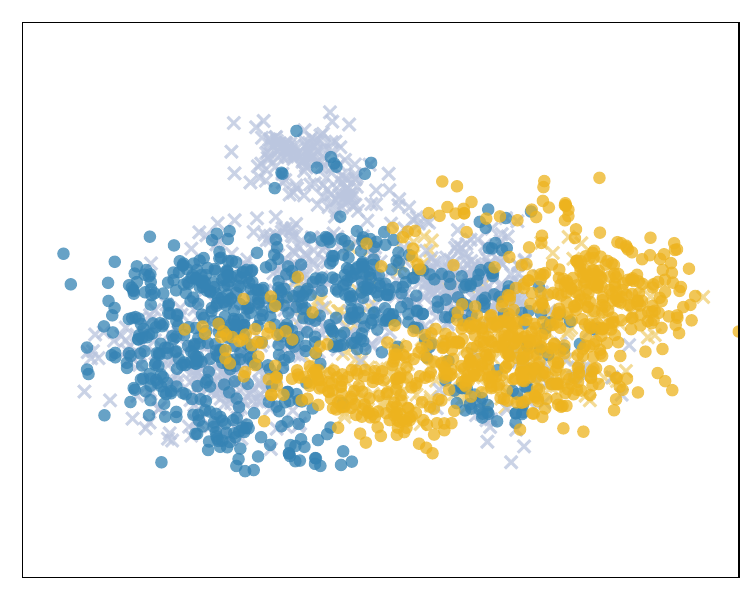} 
        \caption{Represented by \(g(\cdot)\).}
        \label{fig:span_scatter_base}
    \end{subfigure}
    \hspace{4mm}
    \begin{subfigure}[b]{0.20\textwidth}
        \centering
        \includegraphics[width=\textwidth]{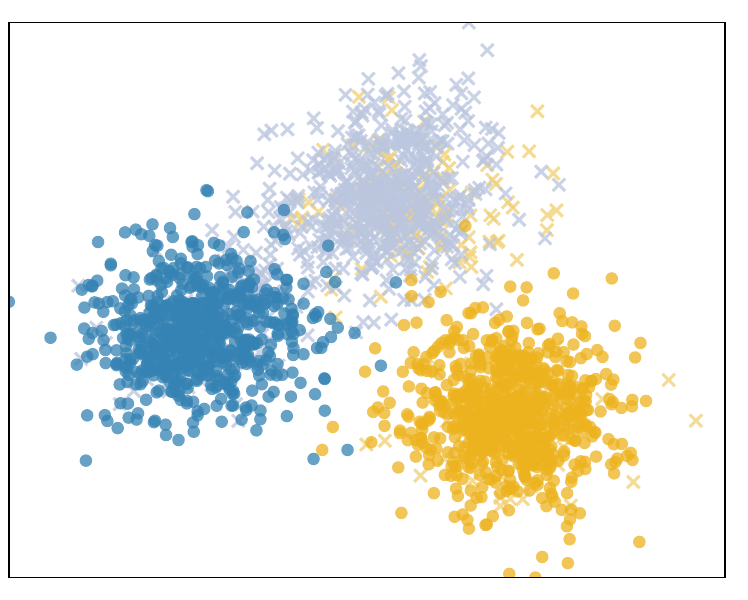} 
        \caption{Represented by \(h(\cdot)\).}
        \label{fig:span_scatter_lora}
    \end{subfigure}
    \vspace{-2mm}
    \caption{\textbf{T-SNE visualization} of span representations produced by backbone \(g(\cdot)\) vs.\ span encoder \(h(\cdot)\).}
    \label{fig:span_scatter}
\end{figure}

\noindent \textbf{Privacy Prototype Generation.\;}
Privacy criteria exhibit inherent domain heterogeneity. Beyond generic PII, each domain adheres to distinct sensitivity standards, exemplified by disease severity in the medical field and crime classification in the legal domain.
For each domain \(k \in \mathcal{K}\), we collect annotated privacy spans from the training corpus \(\mathcal{D}\) into \(\mathcal{S}_k = \{s_{k,i}\}_{i=1}^{N_k}.\)
Ideally, these diverse spans could be summarized into compact prototypes via clustering.
However, as shown in Figure~\ref{fig:span_scatter}, raw representations from pre-trained encoders are highly anisotropic and unstructured. Specifically, privacy and non-privacy spans across domains are entangled in the embedding space and  
lacking discriminative boundaries for effective clustering.

To rectify this, we employ \textbf{multi-domain contrastive learning} to reshape the embedding manifold. Our objective is to enhance discriminability by encouraging intra-domain compactness among privacy spans while enforcing separability from non-privacy contexts and out-of-domain privacy spans.
We instantiate a trainable span encoder $h(\cdot)$ by augmenting a frozen backbone $g(\cdot)$ with lightweight adapter layers.
For an anchor privacy span $s_{k,i} \in \mathcal{S}_k$ with embedding $\mathbf{z}_{k,i} = h(s_{k,i})$, we construct the contrastive objectives based on:
\begin{itemize}[itemsep=1pt, topsep=1pt, leftmargin=10pt]
\item \textit{Positive set} $\mathcal{A}_{k,i}$ consists of other privacy spans within the same domain $k$ (i.e., $\mathcal{S}_k \setminus \{s_{k,i}\}$).
\vspace{-2mm}
\item \textit{Negative set} $\mathcal{N}_{k,i}$ comprises privacy spans from disparate domains ($\mathcal{S}_{j}, j \neq k$) and all non-privacy spans (i.e., $\mathcal{U} \setminus \mathcal{S}$), which serve as hard negatives to enforce semantic discrimination.
\end{itemize}
We minimize the Multi-positive InfoNCE loss to learn discriminative span representations:
\begin{equation}
  \label{eq:infonceloss}
  \mathcal{L}_{\text{CTR}} = -\log
  \frac{\sum_{\mathbf{a} \in \mathcal{A}_{k,i}} \exp\big(\cos(\mathbf{z}_{k,i}, \mathbf{a}) / \tau_1\big)}
       {\sum_{\mathbf{g} \in \mathcal{G}_{k,i}} \exp\big(\cos(\mathbf{z}_{k,i}, \mathbf{g}) / \tau_1\big)},
\end{equation}
where $\mathcal{G}_{k,i} = \mathcal{A}_{k,i} \cup \mathcal{N}_{k,i}$ encompasses all positive and negative samples, and $\tau_1$ is the temperature.

Post-contrastive tuning, the encoder \(h(\cdot)\) yields domain-aware representations. We then derive discrete privacy prototypes by applying the FINCH clustering algorithm~\citep{sarfraz2019efficient} to the refined privacy embeddings of each domain \(k\):
\begin{equation}
  \label{eq:kmeans}
  \mathcal{P}_k = \{\mathbf{p}_{k,j}\}_{j=1}^{J_k}
  = \textsc{Finch}\big(\{\mathbf{z}_{k,i}\}_{i=1}^{N_k}\big).
\end{equation}
These prototypes serve as a structured abstraction of diverse in-domain privacy expressions, providing the foundation for our subsequent steps.

\vspace{0.5mm}
\noindent \textbf{Preference Construction.}
To align the rewriter with both privacy protection and domain consistency, we propose a preference learning framework guided by the learned domain prototypes. 
Rather than training an explicit reward model, we construct a preference dataset with a composite reward function.
For an input $x$ with annotated privacy spans $\mathcal{S}_k^x$, we first generate a set of candidate rewrites $\mathcal{Y}(x)$ using a reference model $\pi_{\text{ref}}$ (detailed in Appendix~\ref{app:template_pref_cons}). We then evaluate each candidate $y \in \mathcal{Y}(x)$ using two competing criteria. Let $\mathbf{z}_{k,i}^{x} = h(s_{k,i}^{x})$ and $\mathbf{z}_{k,i}^{y} = h(s_{k,i}^{y})$ denote the embeddings of the $i$-th original and rewritten privacy spans, respectively.

\noindent \textit{Semantic Obfuscation (Privacy).}
To ensure privacy, the rewritten spans should be semantically distant from the original sensitive information. We define the privacy reward $r_{\text{priv}}(y)$ as the negative cosine similarity between the embeddings of the rewritten spans and the original spans:
\begin{equation}
\label{eq:reward_priv}
    r_{\text{priv}}(y) = 1 - \frac{1}{N_k^x} \sum\nolimits_{i=1}^{N_k^x} \cos\big(\mathbf{z}_{k,i}^{y}, \mathbf{z}_{k,i}^{x}\big).
\end{equation}

\noindent \textit{Domain Fidelity (Utility).}
To preserve domain style, rewritten spans must align with the latent structure of the target domain. We quantify utility $r_{\text{util}}(y)$ by measuring the proximity of each rewritten span to its nearest domain prototype:
\begin{equation}
\label{eq:reward_util}
    r_{\text{util}}(y) = \frac{1}{N_k^x} \sum\nolimits_{i=1}^{N_k^x} \max_{j} \cos\big(\mathbf{z}_{k,i}^{y}, \mathbf{p}_{k,j}\big).
\end{equation}
Here, the $\max$ operator automatically selects the most relevant prototype for each span.

The final score is defined as 
\begin{equation}
  \label{eq:reward_model}
  r(y) = (1-\alpha)\, r_{\text{priv}}(y) + \alpha\, r_{\text{util}}(y),
\end{equation}
where \(\alpha \in [0,1]\) is a trade-off hyperparameter. We leverage the composite reward to automatically construct a preference dataset $\mathcal{D}_{\text{pref}}$. For each input $x$, we identify the candidate with the highest score $r(y)$ as the preferred response $y_w$ and the one with the lowest score as the dispreferred response $y_l$.

\noindent \textbf{Preference Alignment.} 
We fine-tune the rewriter $\pi_\theta$ using DPO, which optimizes the policy to align these synthesized preferences while constraining deviation from the reference model $\pi_{\text{ref}}$:
\begin{equation}
  \label{eq:dpoloss}
  \mathcal{L}_{\text{DPO}}
  = - \log \sigma\Big(
      \beta \log \tfrac{\pi_\theta(y_{w}|x)}{\pi_{\text{ref}}(y_{w}| x)}
      - \beta \log \tfrac{\pi_\theta(y_{l}|x)}{\pi_{\text{ref}}(y_{l}| x)} \Big),
\end{equation}
where $\beta$ regulates the KL-divergence penalty. By focusing optimization on privacy spans, this mechanism effectively achieves a granular balance between semantic privacy and domain style.

\vspace{-1mm}
\subsection{Online Inference Phase} \label{sec:online}
\vspace{-1mm}
We deploy the privacy rewriter as a plug-and-play module on the client side that adaptively rewrites queries prior to transmission to the cloud LLM.

\vspace{0.5mm}
\noindent\textbf{Text Segmentation.}
Given a user query $x$, we first apply a rule-based segmenter, $\text{TextChunker}(\cdot)$, to decompose $x$ into a sequence of semantically coherent spans $\{a_i\}_{i=1}^{M}$ based on punctuation and linguistic cues (detailed in Appendix~\ref{app:TextChunker}).

\vspace{0.5mm}
\noindent\textbf{Privacy Span Localization.}
As the domain of user input is unknown at inference, we use the learned privacy prototypes to jointly infer the global domain and localize sensitive spans. 
Specifically, each span $a_i$ is encoded as $\mathbf{z}_i = h(a_i)$. 
For a candidate domain $k$ with prototype set $\mathcal{P}_k$, we define the affinity of span $a_i$ to domain $k$ by its maximum cosine similarity to the prototypes in $\mathcal{P}_k$:
\begin{equation}
  d_{k,i} = \max\nolimits_{1 \le j \le J_k} \cos(\mathbf{z}_i, \mathbf{p}_{k,j}).
\end{equation}
We then aggregate these span-level affinities across the query and infer the global domain $\hat{k}$ as the one with the highest average affinity:
\begin{equation}
\hat{k} = \arg\max_{k \in \mathcal{K}} \frac{1}{M} \sum\nolimits_{i=1}^{M} d_{k,i}.
\end{equation}
Given the inferred domain $\hat{k}$, we identify a span $a_i$ as private if its affinity $d_{\hat{k},i}$ exceeds a domain-specific threshold $\gamma_{\hat{k}}$. The resulting privacy span set is denoted by $\hat{\mathcal{S}}_{\hat{k}}^x$, while the remaining spans are treated as non-sensitive context. 

To avoid manually setting $\gamma_{\hat{k}}$, we determine it adaptively using the maximum between-cluster variance criterion~\citep{4310076}. We first sort the affinities as $d_{\hat{k},(1)} \leq d_{\hat{k},(2)} \leq \cdots \leq d_{\hat{k},(M)}$. 
For a split point $t$, let $\mu_{l}=\frac{1}{t}\sum_{i=1}^t d_{\hat{k},(i)}$, $\mu_{r}=\frac{1}{M-t}\sum_{i=t+1}^M d_{\hat{k},(i)}$, and $\mu=\frac{1}{M}\sum_{i=1}^M d_{\hat{k},(i)}$. 
The corresponding between-cluster variance is
\begin{equation}
\sigma^2(t)=\frac{t}{M}(\mu_{l}-\mu)^2+\frac{M-t}{M}(\mu_{r}-\mu)^2.
\label{eq:sigmat}
\end{equation}
We select the split point $t^\star=\arg\max_t \sigma^2(t)$ and set the threshold to the midpoint between the two adjacent sorted affinities:
\begin{equation}
\gamma_{\hat{k}}=\big(d_{\hat{k},(t^\star)}+d_{\hat{k},(t^\star+1)}\big)/2.
\end{equation}

\vspace{0.5mm}
\noindent\textbf{DP Span-level Rewriting.}
Given an input query $x$ and detected privacy spans $\hat{\mathcal S}(x)$, our rewriter copies all non-private tokens verbatim and regenerates \emph{only} tokens within $\hat{\mathcal S}(x)$. The privacy guarantee is therefore \emph{span-restricted}: it protects changes within $\hat{\mathcal S}(x)$ while preserving the remaining context exactly. Prompt templates are in Appendix~\ref{app:template_rewrite}.

At each decoding step $\ell$ for a private token, the rewriter outputs a logit vector $u_{\ell}\in\mathbb{R}^{|\mathcal V|}$ over the vocabulary $\mathcal V$. 
 We bound the utility sensitivity by coordinate-wise clipping,
\(\bar u_{\ell}=\mathrm{clip}(u_{\ell};R_1,R_2),\)
and sample the next token with a temperature-softmax distribution, which instantiates the Exponential Mechanism~\citep{mcsherry2007mechanism} with utility $\bar u_{\ell}(\cdot)$:
\begin{equation}
\label{eq:softmax-sampling-rewrite}
\Pr[w_{\ell}=w | x] =
\frac{\exp\!\big(\bar u_{\ell}(w)/\tau_2\big)}
{\sum_{w' \in \mathcal V}\exp\!\big(\bar u_{\ell}(w')/\tau_2\big)}.
\end{equation}
Under clipping, each coordinate satisfies $\bar u_{\ell}(w)\in[R_1,R_2]$, hence the per-step utility sensitivity is bounded by
\(\Delta u \triangleq \max_{x\sim x'} \max_{w\in\mathcal V}\bigl|\bar u_{\ell}(w;x)-\bar u_{\ell}(w;x')\bigr|
\le (R_2-R_1),\)
which yields a per-token privacy cost:
\begin{equation}
\label{eq:eps-token}
\epsilon_{\text{token}}=\frac{2\Delta u}{\tau_2}\le\frac{2(R_2-R_1)}{\tau_2}.
\end{equation}

Let $n_{\text{sp}}(x)$ be the number of regenerated tokens within $\hat{\mathcal S}(x)$, and let $n_{\text{sp}}^{\max}$ be a global upper bound used for budgeting. By sequential composition over the \emph{regenerated} tokens, the rewriting satisfies
\begin{equation}
\label{eq:eps-text}
\epsilon_{\text{text}}(x)\le n_{\text{sp}}(x)\cdot \epsilon_{\text{token}}
\le n_{\text{sp}}^{\max}\cdot \epsilon_{\text{token}}.
\end{equation}
We treat $\epsilon_{\text{text}}$ as the \emph{total privacy budget for all regenerated tokens in a query}, and set the sampling temperature by inverting~\eqref{eq:eps-token}--\eqref{eq:eps-text}:
\begin{equation}
\label{eq:tau2-invert}
\tau_2 \triangleq \frac{2(R_2-R_1)\,n_{\text{sp}}^{\max}}{\epsilon_{\text{text}}}.
\end{equation}
This calibration ensures that the overall privacy loss scales with the amount of \emph{sensitive} text being regenerated, rather than the full query length. Moreover, since $n_{\text{sp}}(x)\le n_{\text{sp}}^{\max}$, the realized privacy loss is often strictly smaller than the budgeted $\epsilon_{\text{text}}$.

\begin{theorem}
\label{thm:span-dp-rewrite}
Condition on the detected span set $\hat{\mathcal S}(x)$ and consider adjacent inputs $x\sim x'$ that differ only within token positions covered by $\hat{\mathcal S}(\cdot)$ (tokens outside $\hat{\mathcal S}(\cdot)$ are identical and copied deterministically). If private tokens are sampled by~\eqref{eq:softmax-sampling-rewrite} with clipping bounds $[R_1,R_2]$ and temperature $\tau_2$ set as in~\eqref{eq:tau2-invert}, then the randomized rewriting stage satisfies $\epsilon_{\text{text}}$-DP with $\epsilon_{\text{text}}=n_{\text{sp}}^{\max}\epsilon_{\text{token}}$.
\end{theorem}
\vspace{-1mm}
\section{Experiments}\label{sec:Experiments}
\vspace{-1mm}
\subsection{Experimental Setup} 
\vspace{-1mm}
\noindent\textbf{Datasets.}
We evaluate on two established datasets: Pri-DDXPlus~\citep{zeng-etal-2025-privacyrestore} (Medical) and Pri-SLJA~\citep{zeng-etal-2025-privacyrestore} (Legal). To assess domain perception and adaptability, we further construct a composite benchmark, Pri-Mixture, by aggregating samples from these base domains. Detailed statistics are provided in Appendix~\ref{app:datasets}.
 
\vspace{0.5mm}
\noindent \textbf{Training Protocol.} 
To demonstrate the framework's capability for automatic domain-aware privacy, \textit{we train the model on Pri-Mixture to learn discriminative domain features} and evaluate it on Pri-DDXPlus, Pri-SLJA, and Pri-Mixture. This protocol challenges the model to dynamically adapt to varying contexts without explicit task indicators. Unless otherwise specified, all results report the performance of this unified model.

\vspace{0.5mm}
\noindent\textbf{Baselines.}
We compare against No Rewriting (upper bound), DP-Paraphrase~\citep{mattern-etal-2022-limits}, DP-Prompt~\citep{utpala-etal-2023-locally}, PrivacyRestore~\citep{zeng-etal-2025-privacyrestore}, and DP-MLM~\citep{meisenbacher-etal-2024-dp}.
To isolate localization effects, we introduce two analysis settings:
(1) Oracle variants (DP-MLM$_{\text{oracle}}$, DAMPER$_{\text{oracle}}$), using ground-truth privacy spans to evaluate the rewriter in isolation.
(2) DP-MLM$_{\text{auto}}$, a hybrid baseline that applies the DP-MLM rewriter specifically to spans detected by our prototype-guided localizer.

\vspace{0.5mm}
\noindent \textbf{Metrics.}
We evaluate three dimensions: 
downstream utility via Accuracy (ACC), semantic consistency via BERTScore (BS), and generation quality via LLM-Judge (LLM-J)~\citep{zheng2023judging, zeng-etal-2025-privacyrestore}. Definitions and evaluation prompt templates are provided in Appendix~\ref{app:metrics} and~\ref{app:template}.

\vspace{0.5mm}
\noindent\textbf{Implementation Details.} 
We utilize RoBERTa-base~\citep{Liu2019RoBERTaAR} as the backbone for the span encoder and Qwen2.5-1.5B-Instruct~\cite{qwen2.5} as reference and policy models for preference learning. The cloud LLM is instantiated with Qwen2.5-7B-Instruct~\citep{qwen2.5,qwen2}.
Default hyperparameters are configured as $\tau_1=0.1$, $\beta=0.1$, and $\alpha=0.3$. We adopt the same privacy hyperparameter $\epsilon$ from PrivacyRestore~\citep{zeng-etal-2025-privacyrestore}.
The $\epsilon_\text{token}$ of DAMPER is 2.88 on Pri-DDXPlus and 1.44 on Pri-SLJA.
See Appendix~\ref{app:implementationdetails} for $\epsilon_\text{text}$, $\epsilon_\text{token}$ and $\epsilon$ details.

\vspace{-1mm}
\subsection{Performance Comparison}
\vspace{-1mm}
\begin{table*}[t] 
\centering
\small
\renewcommand{\arraystretch}{1.25}
\fontsize{8}{7.8}\selectfont
\setlength{\tabcolsep}{1pt}
\begin{tabularx}{\linewidth}{
l||*{3}{>{\centering\arraybackslash}X}|
   *{3}{>{\centering\arraybackslash}X}|
   *{3}{>{\centering\arraybackslash}X}}
\noalign{\hrule height 1pt}
\rowcolor{gray!25}
& \multicolumn{3}{c|}{\textbf{Pri-DDXPlus}} & \multicolumn{3}{c|}{\textbf{Pri-SLJA}} & \multicolumn{3}{c}{\textbf{Pri-Mixture}} \\
\cline{2-4} \cline{5-7} \cline{8-10} 
\rowcolor{gray!25}
\multicolumn{1}{c||}{\multirow{-2}{*}{Methods}} & ACC ${\uparrow}$ & BS ${\uparrow}$ & LLM-J ${\uparrow}$ & ACC ${\uparrow}$ & BS ${\uparrow}$ & LLM-J ${\uparrow}$ & ACC ${\uparrow}$ & BS ${\uparrow}$ & LLM-J ${\uparrow}$ \\
\hline\hline
No Rewriting & 87.60$_{\pm 0.07}$ & 1.00$_{\pm 0.00}$ & 5.60$_{\pm 0.14}$ & 89.79$_{\pm 0.28}$ & 1.00$_{\pm 0.00}$ & 6.89$_{\pm 0.08}$ & 88.45$_{\pm 0.14}$ & 1.00$_{\pm 0.00}$ & 6.04$_{\pm 0.12}$ \\
\hline\hline
\rowcolor{gray!10}
DP-Paraphrase & 45.35$_{\pm 0.71}$ & 0.44$_{\pm 0.00}$ & 1.66$_{\pm 0.02}$ & 46.86$_{\pm 0.97}$ & 0.41$_{\pm 0.00}$ & 1.94$_{\pm 0.04}$ & 43.06$_{\pm 0.33}$ & 0.43$_{\pm 0.00}$ & 1.76$_{\pm 0.03}$ \\
DP-Prompt & 29.56$_{\pm 0.57}$ & 0.28$_{\pm 0.00}$ & 1.68$_{\pm 0.01}$ & 23.97$_{\pm 0.83}$ & 0.23$_{\pm 0.00}$ & 2.14$_{\pm 0.16}$ & 27.09$_{\pm 0.90}$ & 0.27$_{\pm 0.00}$ & 1.84$_{\pm 0.06}$ \\
\rowcolor{gray!10}
DP-MLM & 33.12$_{\pm 0.39}$ & 0.39$_{\pm 0.00}$ & 1.70$_{\pm 0.06}$ & 25.93$_{\pm 2.91}$ & 0.28$_{\pm 0.00}$ & 2.22$_{\pm 0.04}$ & 28.93$_{\pm 1.25}$ & 0.35$_{\pm 0.00}$ & 1.88$_{\pm 0.03}$ \\
DP-MLM$_{\text{oracle}}$ & 50.94$_{\pm 0.88}$ & \underline{0.59}$_{\pm 0.00}$ & 3.61$_{\pm 0.10}$ & \underline{78.78}$_{\pm 0.69}$ & \textbf{0.73}$_{\pm 0.00}$ & 4.81$_{\pm 0.02}$ & 59.60$_{\pm 0.91}$ & \underline{0.64}$_{\pm 0.00}$ & 4.02$_{\pm 0.06}$ \\
\rowcolor{gray!10}
PrivacyRestore & \underline{75.82}$_{\pm 0.17}$ & - & \underline{4.32}$_{\pm 0.01}$ & 78.38$_{\pm 1.02}$ & - & \underline{5.26}$_{\pm 0.03}$ & \underline{76.35}$_{\pm 0.56}$ & - & \underline{4.64}$_{\pm 0.02}$  \\
\hline
\rowcolor{lightblue}
DAMPER  & \textbf{78.13}$_{\pm 1.13}$ & \textbf{0.68}$_{\pm 0.00}$ & \textbf{4.76}$_{\pm 0.21}$ & \textbf{82.68}$_{\pm 0.86}$ & \underline{0.66}$_{\pm 0.00}$ & \textbf{5.37}$_{\pm 0.04}$ & \textbf{78.29}$_{\pm 0.59}$ & \textbf{0.67}$_{\pm 0.00}$ & \textbf{4.97}$_{\pm 0.09}$  \\

\noalign{\hrule height 0.8pt} 
\end{tabularx}
\vspace{-2mm}
\caption{Performance comparison on three datasets. ACC is reported in \%, where ``-'' denotes unavailable results. Results are averaged over 3 runs. The \textbf{best} and \underline{second} are marked.}
\vspace{-2mm}
\label{tab:main_experiment}
\end{table*}

\begin{figure*}[t]
  \centering
  \begin{subfigure}[b]{0.24\linewidth}
    \centering
    \includegraphics[height=0.118\textheight]{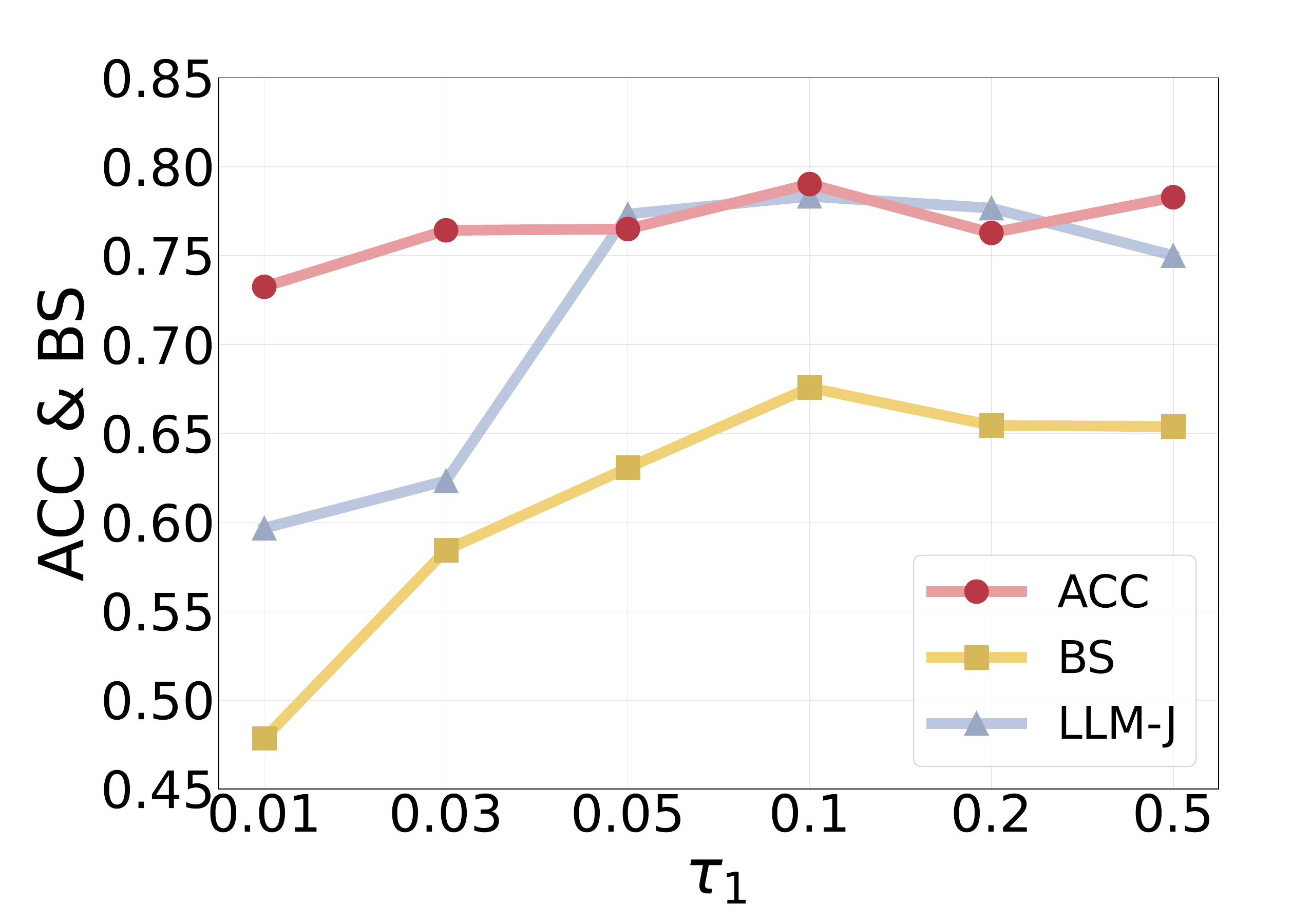}
    \caption{The effect of $\tau_1$.}
    \label{fig:temperature}
  \end{subfigure}
  \hfill
  \begin{subfigure}[b]{0.24\linewidth}
    \centering
    \includegraphics[height=0.118\textheight]{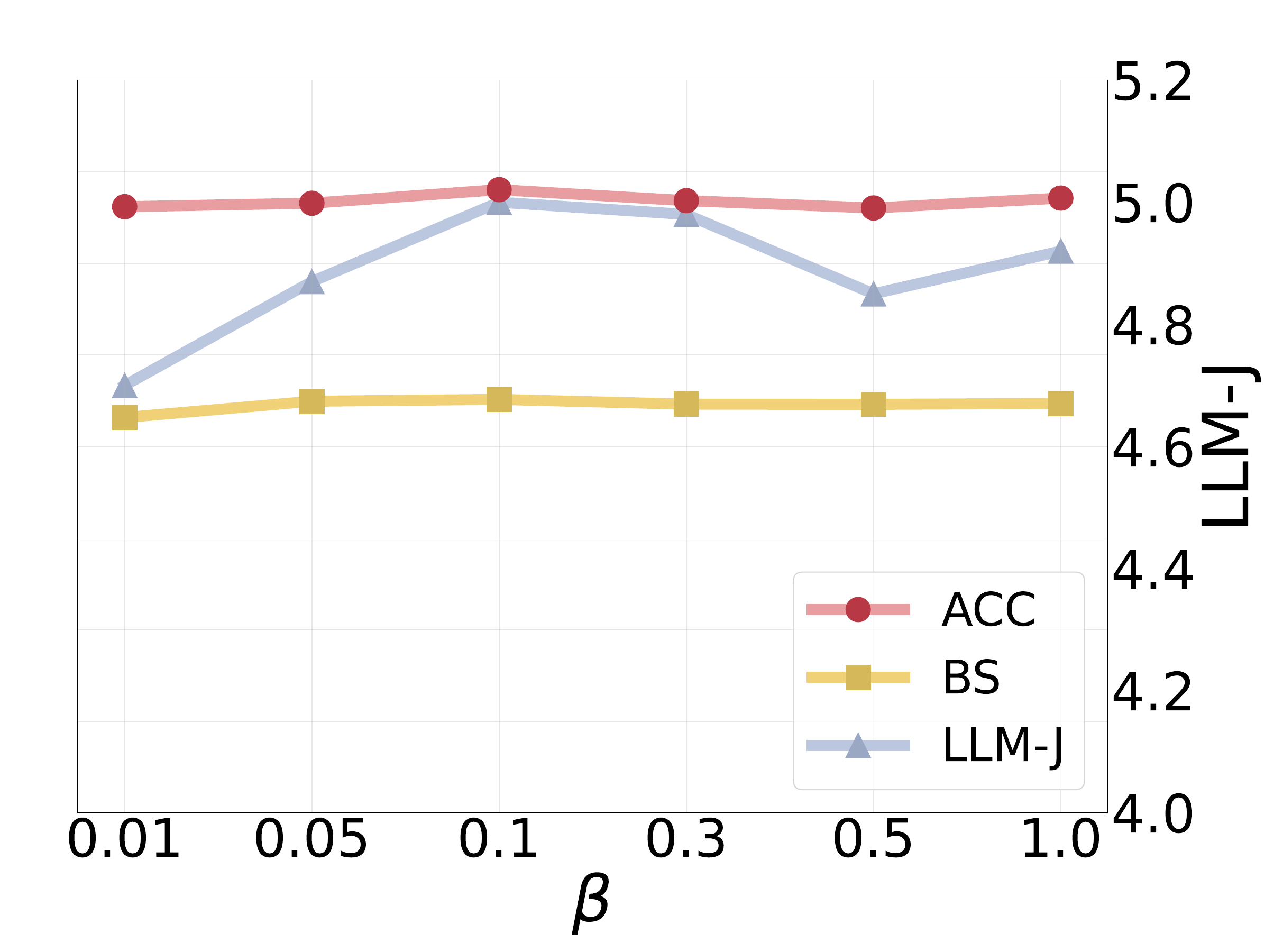}
    \caption{The effect of $\beta$.}
    \label{fig:beta}
  \end{subfigure}
  \hfill
  \begin{subfigure}[b]{0.24\linewidth}
    \centering
    \includegraphics[height=0.118\textheight]{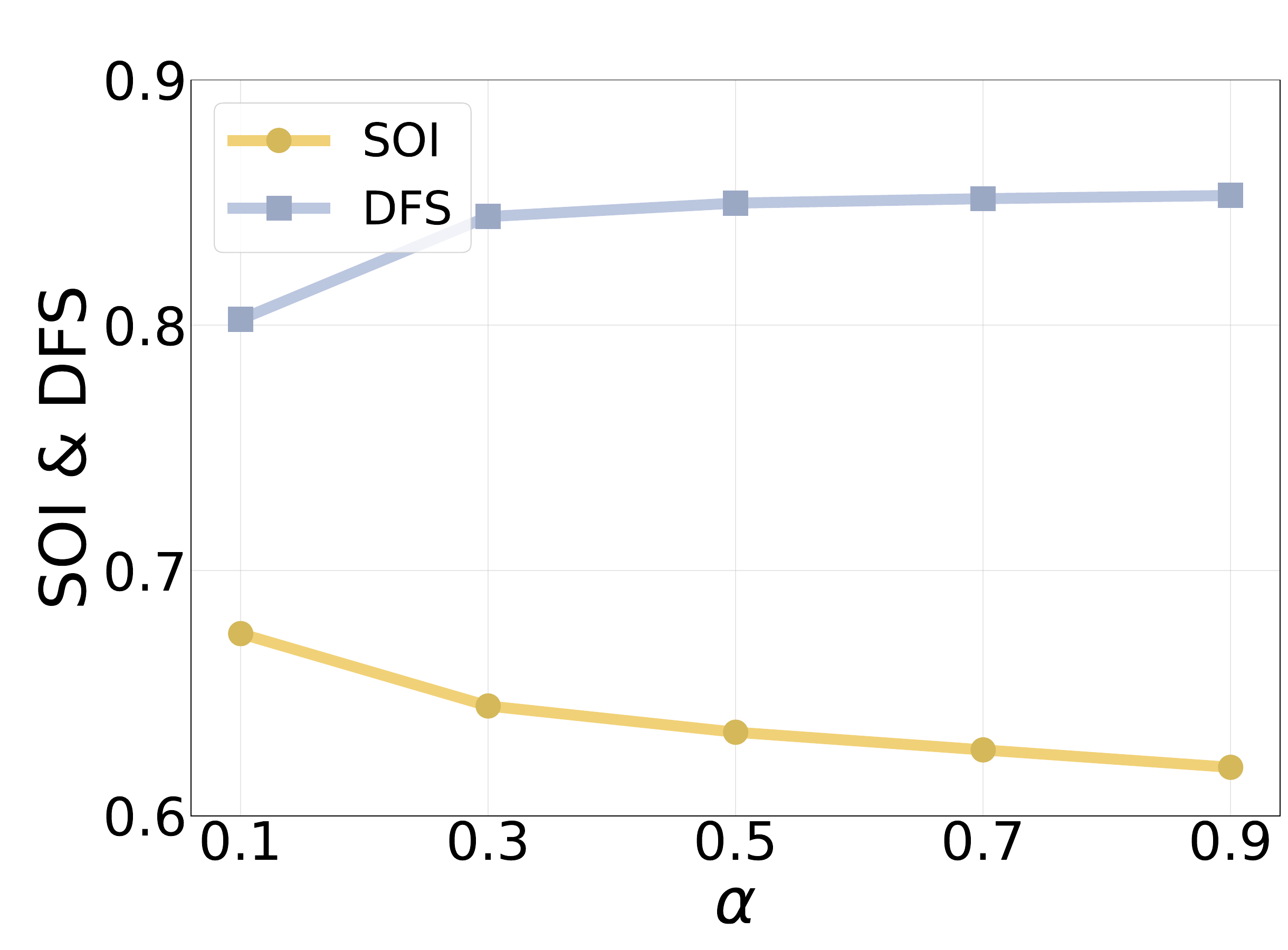}
    \caption{The effect of $\alpha$.}
    \label{fig:alpha_reward}
  \end{subfigure}
  \hfill
  \begin{subfigure}[b]{0.24\linewidth}
    \centering
    \includegraphics[height=0.118\textheight]{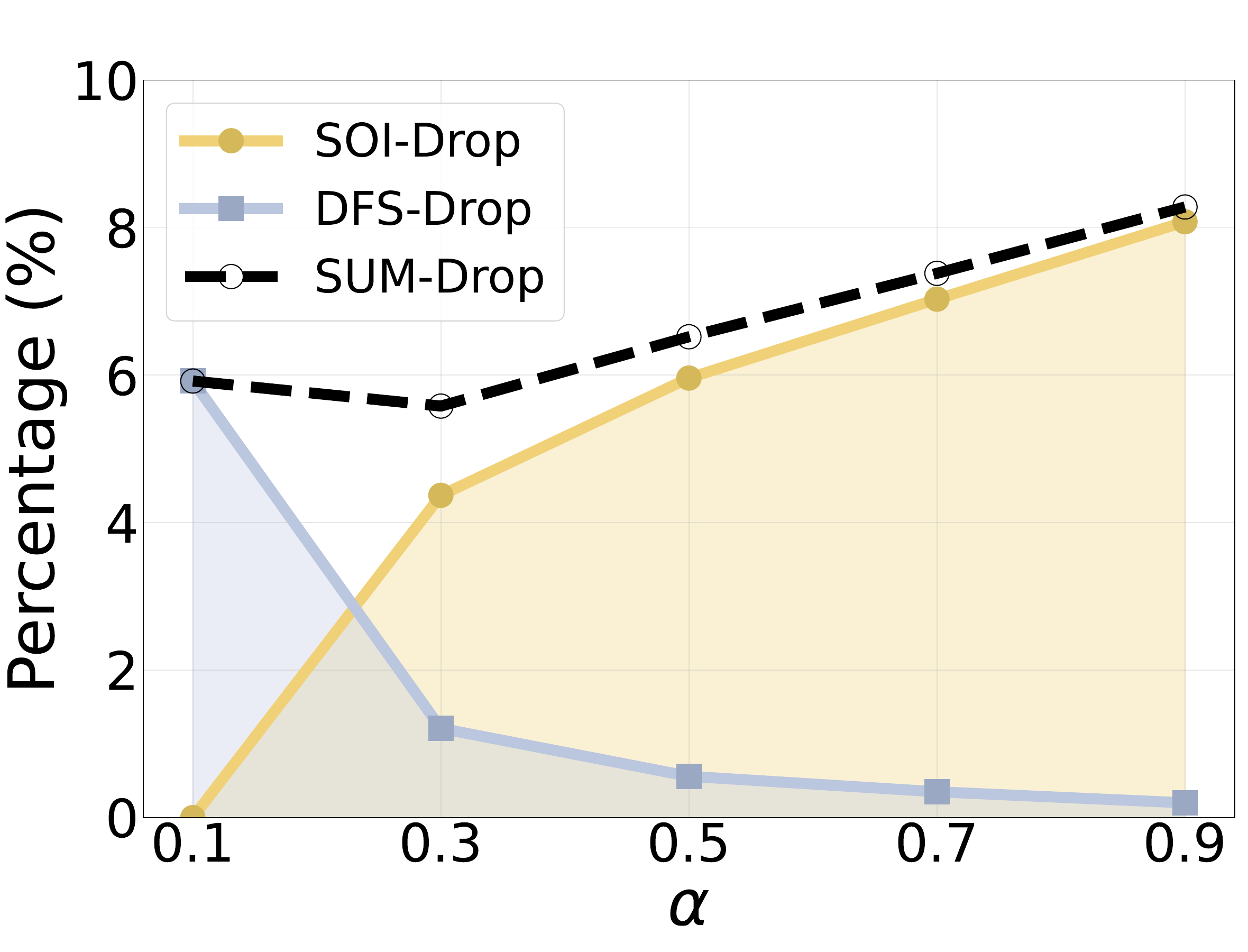}
    \caption{The effect of $\alpha$.}
    \label{fig:alpha_drop}
  \end{subfigure}
  \vspace{-2mm}
  \caption{\textbf{Hyper-parameter sensitivity} of DAPPER on Pri-Mixture. 
  }
  \label{fig:alpha_1x4}
\end{figure*}

Table~\ref{tab:main_experiment} compares DAMPER against six baselines, showing superior or competitive performance across all metrics.
\blackcircnum{1} In terms of downstream utility (ACC), DAMPER establishes a new state of the art. This advantage is pronounced on Pri-Mixture, where baselines suffer from inter-domain interference. DAMPER effectively preserves domain boundaries, confirming that our prototype-driven approach captures discriminative features robust to distractors.
\blackcircnum{2} Regarding semantic consistency (BS), DAMPER leads on Pri-DDXPlus and Pri-Mixture, ranking second only to DP-MLM$_{\text{oracle}}$ on Pri-SLJA. This gap is expected, as the oracle uses ground-truth spans. Despite autonomous inference without masks, DAMPER maintains high semantic fidelity.
\blackcircnum{3} For generation quality (LLM-J), DAMPER generally outperforms baselines. While PrivacyRestore shows a slight edge on Pri-SLJA under high $\epsilon$, it requires expensive retrieval. Conversely, DAMPER achieves comparable utility through efficient one-pass rewriting without post-hoc mechanisms.
More results under varying privacy budgets are provided in Appendix~\ref{app:performance}.

\vspace{-1mm}
\subsection{Diagnostic Analysis} 
\vspace{-1mm}
\noindent\textbf{Hyper-parameter Sensitivity.} 
Figures~\ref{fig:temperature} and~\ref{fig:beta} indicate that performance consistently peaks at $\beta=0.1$ and $\tau_1=0.1$. Regarding the trade-off parameter $\alpha$, Figure~\ref{fig:alpha_reward} confirms that lower values favor privacy (Semantic Obfuscation Index, SOI) while higher values prioritize fidelity (Domain Fidelity Score, DFS). To determine the optimal trade-off, we minimize the cumulative deviation from each metric's optimum (SUM-Drop, detailed in Appendix~\ref{app:metrics}). As shown in Figure~\ref{fig:alpha_drop}, $\alpha=0.3$ achieves the most favorable balance. The optimal values of these parameters are consistent with the default settings used throughout our experiments.

\begin{table}[t]
\centering 
\vspace{-3mm}
\setlength{\tabcolsep}{3pt}
\renewcommand{\arraystretch}{1.25}
\fontsize{8}{7.8}\selectfont
\begin{tabularx}{\columnwidth}{cc||YYY}
\noalign{\hrule height 0.8pt} 
\rowcolor{gray!25} 
& & \multicolumn{3}{c}{\textbf{Pri-Mixture}} \\ 
\cline{3-5}
\rowcolor{gray!25}
\multirow{-2}{*}{$\mathcal{L}_{CTR}$} & \multirow{-2}{*}{$\mathcal{L}_{DPO}$} & ACC ${\uparrow}$ & BS ${\uparrow}$ & LLM-J ${\uparrow}$ \\
\hline\hline
                &                & 36.01 & 0.40 & 2.31 \\
                & \checkmark     & 69.07$_{\scriptscriptstyle \textcolor{blue}{\uparrow 33.06}}$
                                & 0.45$_{\scriptscriptstyle \textcolor{blue}{\uparrow 0.05}}$
                                & 4.21$_{\scriptscriptstyle \textcolor{blue}{\uparrow 1.90}}$ \\
\checkmark      &                & 66.07$_{\scriptscriptstyle \textcolor{blue}{\uparrow 30.06}}$
                                & 0.59$_{\scriptscriptstyle \textcolor{blue}{\uparrow 0.19}}$
                                & 4.12$_{\scriptscriptstyle \textcolor{blue}{\uparrow 1.81}}$ \\
\checkmark      & \checkmark     & \textbf{78.29}$_{\scriptscriptstyle \textcolor{blue}{\uparrow 42.28}}$
                                & \textbf{0.67}$_{\scriptscriptstyle \textcolor{blue}{\uparrow 0.27}}$
                                & \textbf{4.97}$_{\scriptscriptstyle \textcolor{blue}{\uparrow 2.66}}$ \\
\noalign{\hrule height 0.8pt} 
\end{tabularx}
\vspace{-2mm}
\caption{\textbf{Ablation study} on Pri-Mixture.}
\vspace{-2mm}
\label{tab:ablation_modules}
\end{table}

\vspace{0.5mm}
\noindent\textbf{Ablation Study.}  
Table~\ref{tab:ablation_modules} presents the ablation results for $\mathcal{L}_{\text{CTR}}$ and $\mathcal{L}_{\text{DPO}}$ on the Pri-Mixture dataset. Individually, DPO and contrastive learning significantly boost ACC by 33.06\% and 30.06\%, respectively. The latter validates the necessity of mitigating embedding anisotropy, as discussed in Section~\ref{sec:offline}. When combined, performance peaks at 42.28\%, confirming the strong complementary effects of the two modules. Results for other datasets are detailed in Appendix~\ref{app:ablation_full}.

\vspace{0.5mm}
\noindent\textbf{Robustness of Localization.}
Table~\ref{tab:combination_main} assesses the resilience of localization by benchmarking automatic inference (ours) against ground-truth oracle settings. DAMPER exhibits exceptional stability, maintaining performance comparable to $\text{DAMPER}_{\text{oracle}}$ (78.29\% vs.\ 79.67\%). In contrast, $\text{DP-MLM}_{\text{auto}}$ suffers a severe regression against $\text{DP-MLM}_{\text{oracle}}$ (49.63\% vs.\ 59.60\%), exposing a brittle dependency on ground-truth masks. More results of localization are provided in Appendix~\ref{app:combination}.  

\begin{table}[t]
\centering
\vspace{-3mm}
\setlength{\tabcolsep}{3pt}
\renewcommand{\arraystretch}{1.25}
\fontsize{8}{7.8}\selectfont
\begin{tabularx}{\columnwidth}{l||YYY}
\noalign{\hrule height 0.8pt}
\rowcolor{gray!25}
 & \multicolumn{3}{c}{\textbf{Pri-Mixture}} \\
\cline{2-4}
\rowcolor{gray!25}
\multirow{-2}{*}{Methods} & ACC ${\uparrow}$ & BS ${\uparrow}$ & LLM-J ${\uparrow}$\\
\hline\hline
DP-MLM$_{\text{oracle}}$ & 59.60 & 0.64 & 4.02 \\
DP-MLM$_{\text{auto}}$   & 49.63 & 0.56 & 3.50 \\
\hline
DAMPER$_{\text{oracle}}$ & 79.67 & 0.78 & 5.11\\
DAMPER                  & 78.29 & 0.67 & 4.97\\
\noalign{\hrule height 0.8pt}
\end{tabularx}
\vspace{-2mm}
\caption{\textbf{Robustness of localization} on Pri-Mixture under oracle vs.\ automatic (ours) localization.}
\vspace{-2mm}
\label{tab:combination_main}
\end{table}  

We present results on \textbf{TextChunker's robustness} in Appendix~\ref{app:textchunkerrobustness}. We defer additional results on \textbf{single-domain training} to Appendix~\ref{app:singledomain}, with extended \textbf{clustering analyses} available in Appendix~\ref{app:clustering}.

\vspace{-1mm}
\subsection{Generalization to Unseen Privacy Spans}
\label{app:unseenprivacyspans}
\vspace{-1mm}

\begin{figure}[t]
  \centering
  \includegraphics[width=0.95\linewidth]{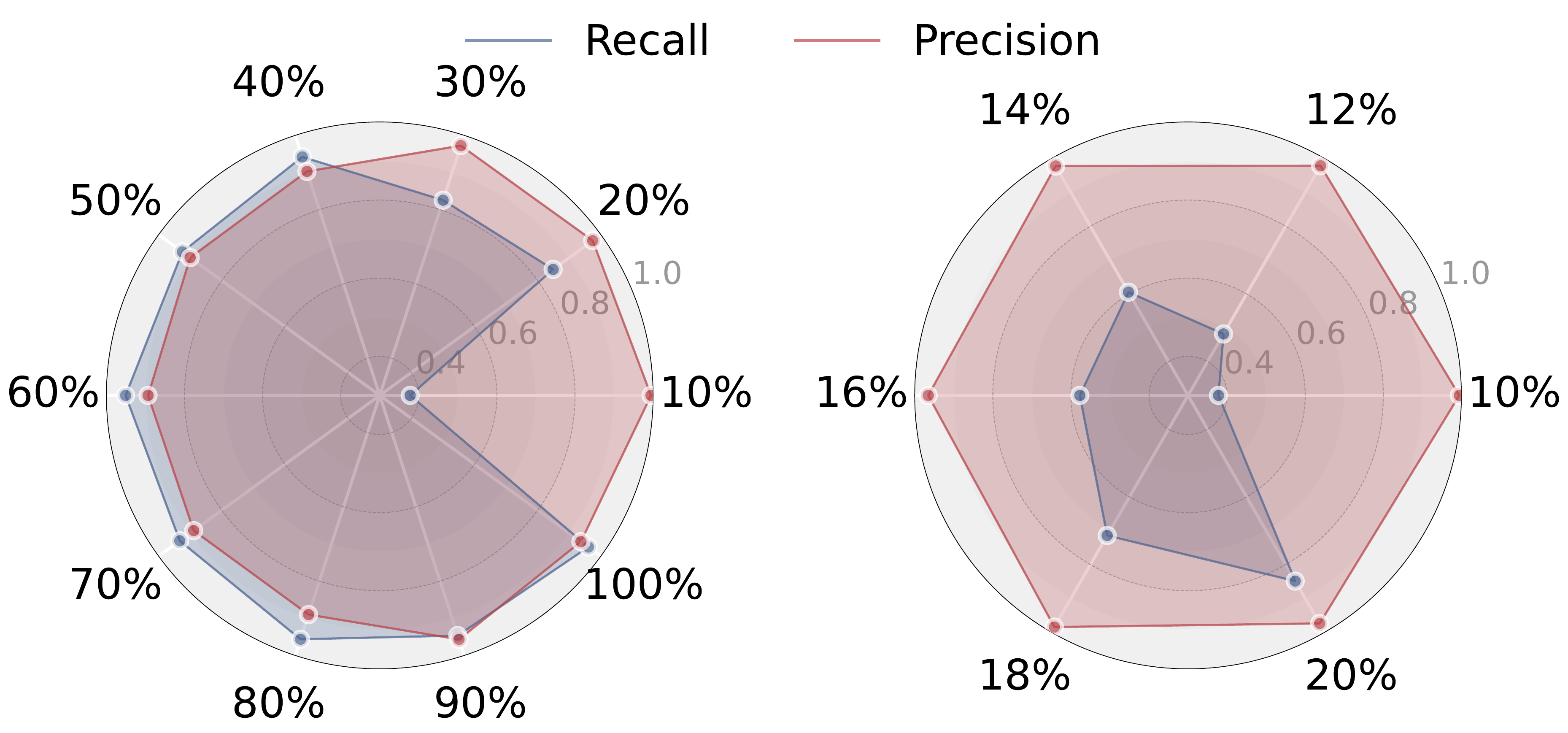}
    \vspace{-2mm}
    \caption{Performance under different \textbf{Top-\%} predefined privacy spans on Pri-DDXPlus.
    }
  \label{fig:generalization}
\end{figure}

Real-world privacy spans typically exhibit a long-tailed distribution~\citep{zeng-etal-2025-privacyrestore}. To assess generalization to unseen or rare instances, we train DAMPER using only the Top-$k$\% most frequent spans and evaluate it on the full test set.
As illustrated in Figure~\ref{fig:generalization}, DAMPER displays exceptional robustness to vocabulary scarcity. It maintains 84.88\% Recall even when restricted to the Top-20\% frequent spans during training, with significant degradation appearing only below this threshold. This confirms that our learned prototypes capture abstract semantic regularities rather than relying on surface-level memorization, enabling effective detection extended to the long tail.

\vspace{-1mm}
\subsection{Privacy Attacks Assessment}
\vspace{-1mm}
To rigorously evaluate the privacy robustness of DAMPER, we conduct two types of adversarial assessments against span-level rewriting methods following the methodology of~\citet{zeng-etal-2025-privacyrestore}.

\begin{figure}[t]
  \centering
  \begin{subfigure}[b]{0.49\linewidth}
    \centering
    \includegraphics[width=\linewidth]{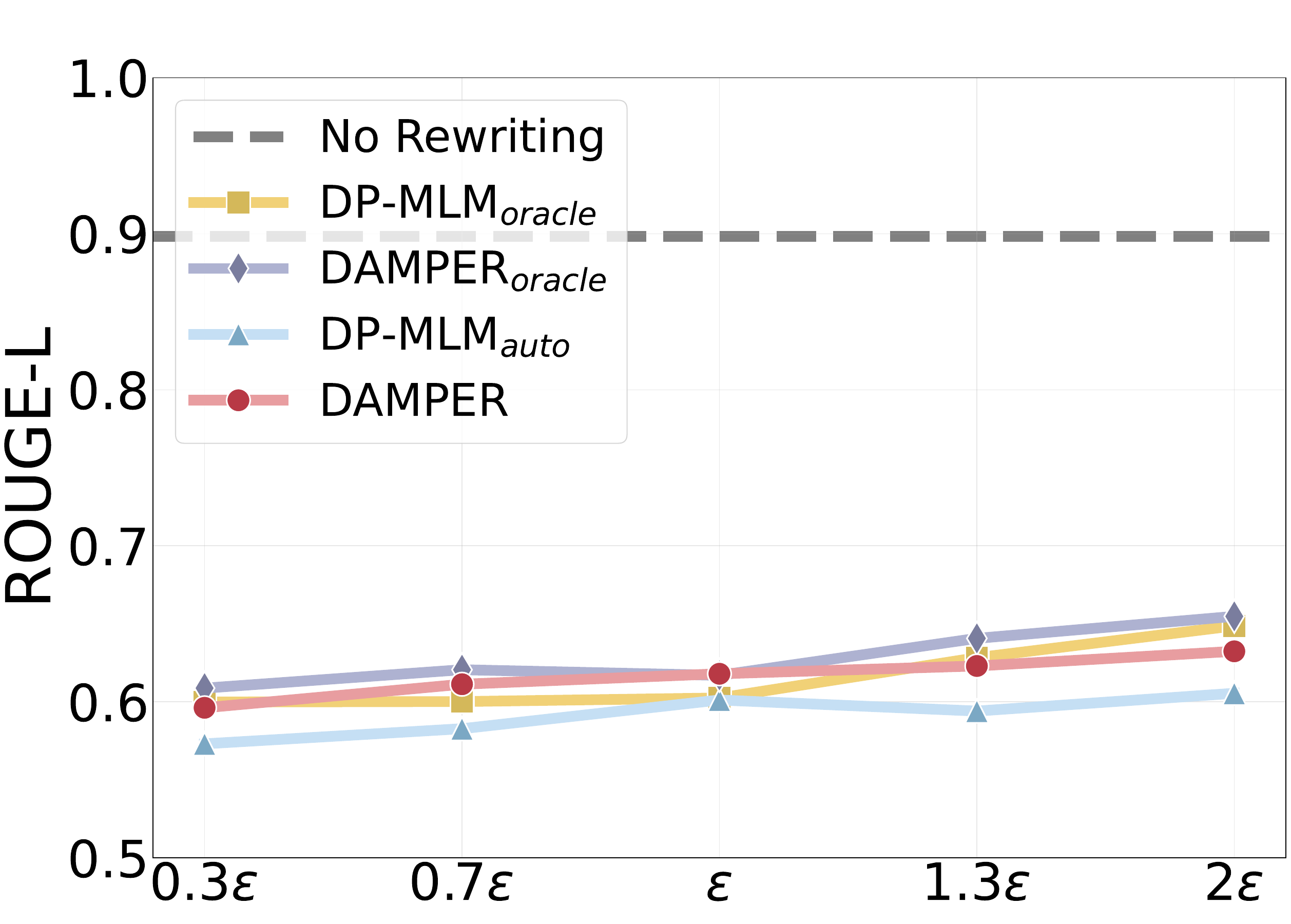}
    \caption{EIA on ROUGE-L.}
    \label{fig:eia_rouge}
  \end{subfigure}
  \hfill
  \begin{subfigure}[b]{0.49\linewidth}
    \centering
    \includegraphics[width=\linewidth]{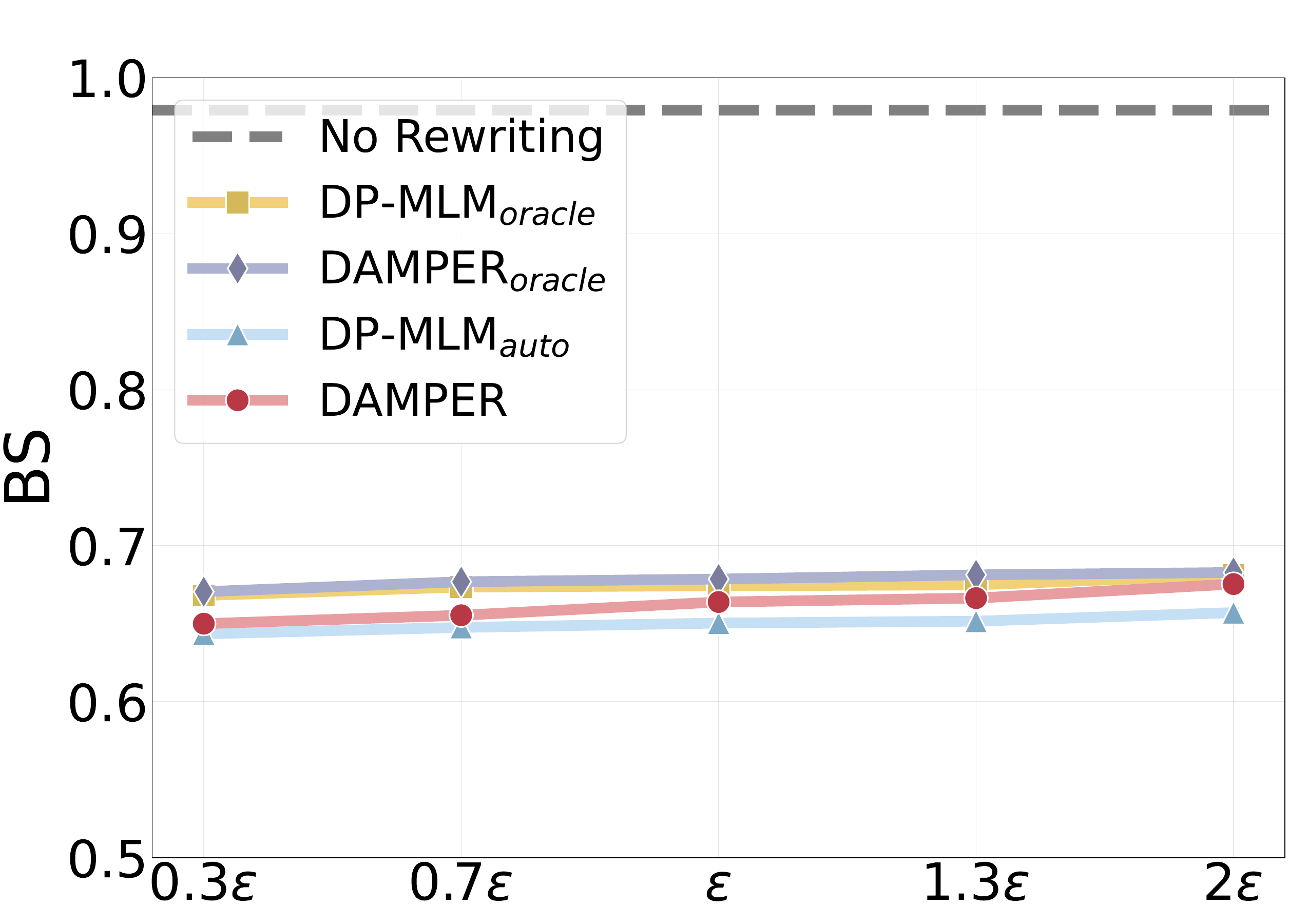}
    \caption{EIA on BS.}
    \label{fig:eia_bs}
  \end{subfigure}

  \vspace{0.2em}

  \begin{subfigure}[b]{0.49\linewidth}
    \centering
    \includegraphics[width=\linewidth]{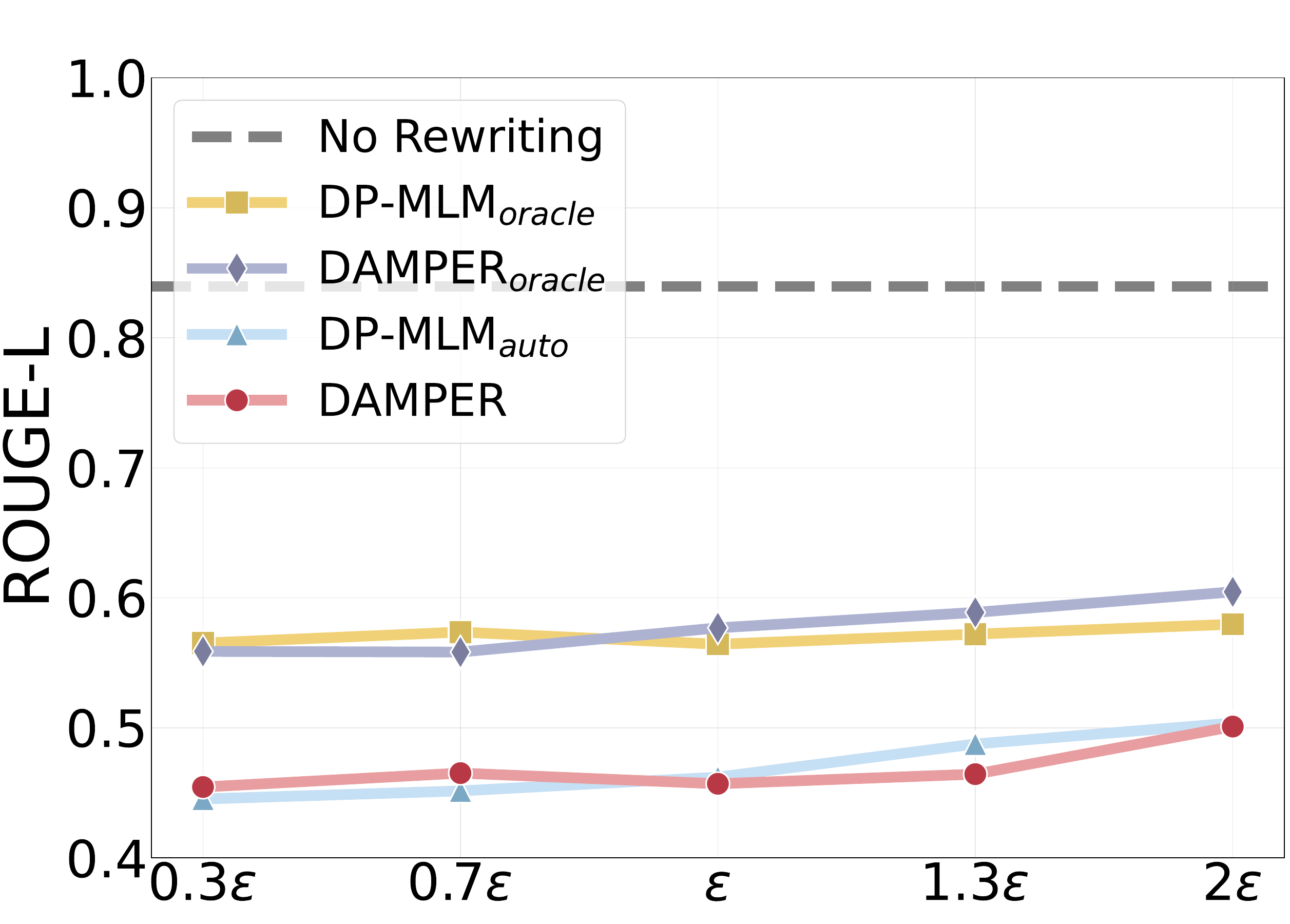}
    \caption{PIA on ROUGE-L.}
    \label{fig:pia_rouge}
  \end{subfigure}
  \hfill
  \begin{subfigure}[b]{0.49\linewidth}
    \centering
    \includegraphics[width=\linewidth]{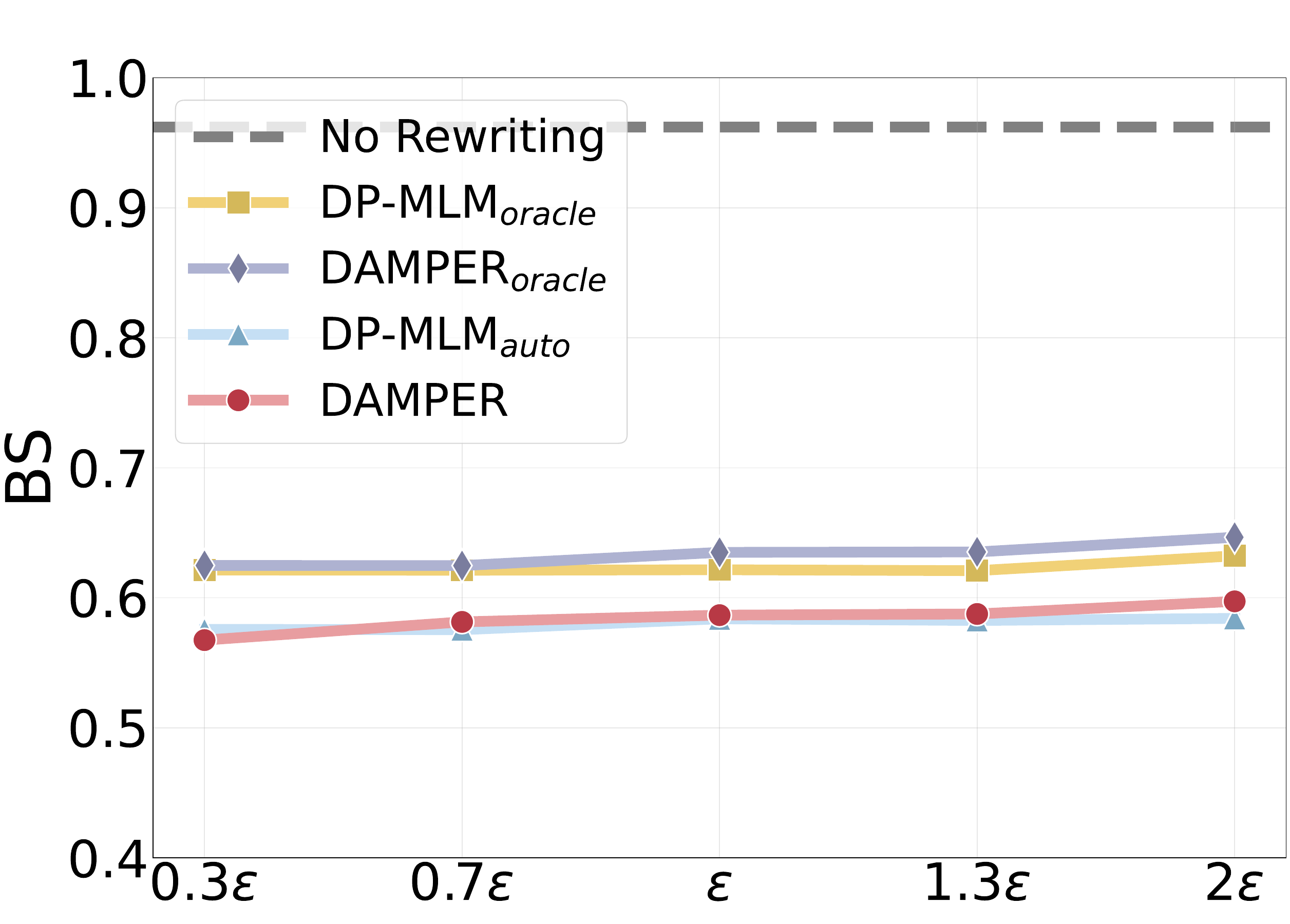}
    \caption{PIA on BS.}
    \label{fig:pia_bs}
  \end{subfigure}
  \vspace{-4mm}
  \caption{Results of privacy attacks on span-level rewriting methods under varying \(\epsilon\) on Pri-SLJA.}
  \label{fig:attack}
\end{figure}

\vspace{0.5mm}
\noindent \textbf{Embedding Inversion Attack (EIA).}
EIA attempts to reconstruct sensitive spans directly from the user's input embeddings by training a dedicated inversion model. We quantify potential leakage using ROUGE-L~\cite{Lin2004ROUGEAP} (see Appendix~\ref{app:metrics}) and BS measured between the reconstructed and ground-truth privacy spans. Empirical results (Figure~\ref{fig:eia_rouge} and ~\ref{fig:eia_bs}) indicate that across varying privacy budgets $\epsilon$, DAMPER maintains a high level of protection comparable to other defense baselines.

\vspace{0.5mm}
\noindent\textbf{Prompt Injection Attack (PIA).} 
PIA aims to elicit the original input from the cloud LLM by injecting adversarial prompts into the rewritten text. We measure the attack success using ROUGE-L and BS between the recovered inputs and the original texts. Across all privacy budgets (Figure~\ref{fig:pia_rouge} and ~\ref{fig:pia_bs}), DAMPER and DP-MLM achieve similar defense capabilities under both oracle and automatic settings. The automatic setting consistently yields stronger privacy protection (i.e., lower leakage) than the oracle setting, as conservative detection often masks additional context. More details for both EIA and PIA are provided in Appendix~\ref{app:attack}.

\vspace{-1mm}
\subsection{Computational Cost}
\vspace{-1mm}
\begin{figure}[t]
  \centering
  \begin{subfigure}[b]{0.48\linewidth}
    \includegraphics[width=\linewidth]{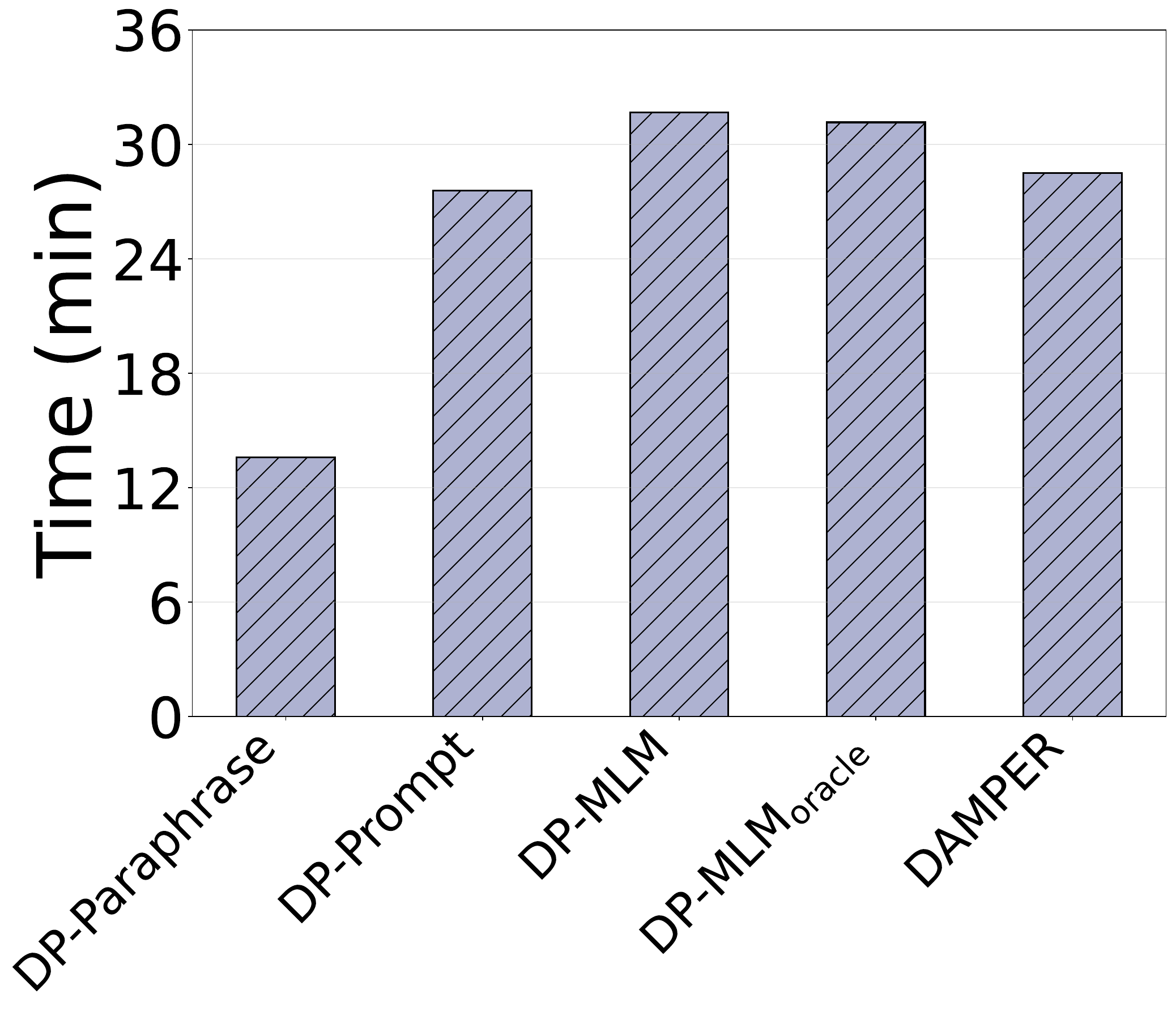}
    \caption{Pri-DDXPlus.} 
    \label{fig:efficiency_ddxplus}
  \end{subfigure}
  \hspace{0.005\linewidth}
  \begin{subfigure}[b]{0.48\linewidth}
    \includegraphics[width=\linewidth]{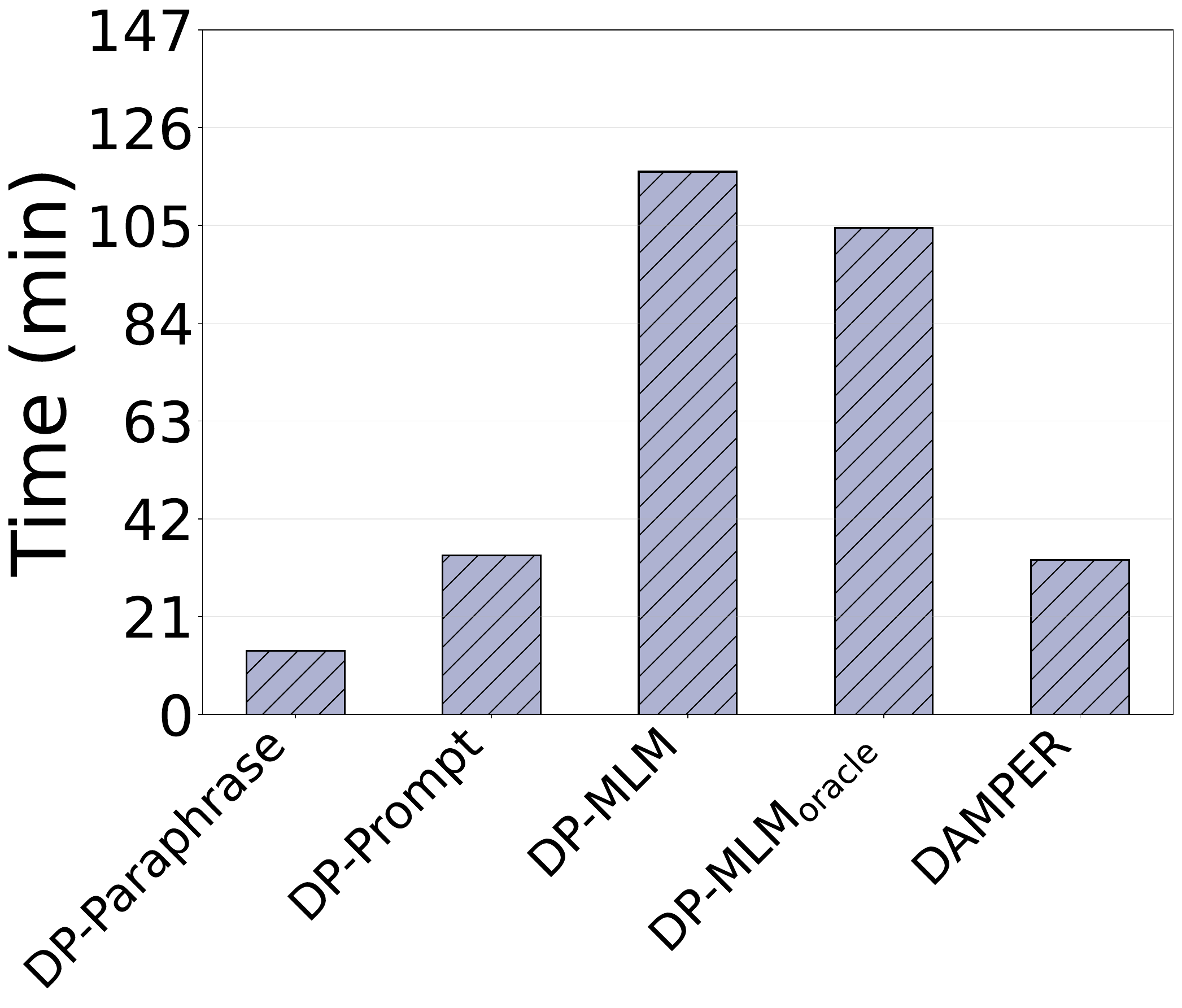}
    \caption{Pri-SLJA.}
    \label{fig:efficiency_slja}
  \end{subfigure}
  \vspace{-2mm}
  \caption{\textbf{Computational cost comparison} of different methods on Pri-DDXPlus and Pri-SLJA}
  \label{fig:efficiency}
\end{figure}

We evaluate DAMPER and four baselines, by performing privacy localization and rewriting on 1013 samples from Pri-DDXPlus and 509 samples from Pri-SLJA, respectively, and report the corresponding time costs. As illustrated in Figure~\ref{fig:efficiency}, across both datasets, DP-MLM and its variants incur the highest time cost, while DP-Paraphrase is the most time-efficient. DAMPER exhibits nearly identical runtime to DP-Prompt. Notably, DP-MLM and its variants exhibit a pronounced increase in runtime as the input text length grows, whereas the other methods remain largely insensitive to text length. This behavior arises from the token-level rewriting strategy adopted by DP-MLM, which processes the input sequentially on a per-token basis.

\vspace{-1mm}
\section{Conclusion}
\vspace{-1mm}
We proposed DAMPER, a client-side privacy rewriting framework for high-stakes, domain-sensitive queries under an untrusted cloud LLM. By inferring the dominant domain and localizing privacy spans with domain-contingent prototypes, DAMPER applies a calibrated DP mechanism only to the sensitive spans, thereby preserving non-sensitive context and task intent. Experiments on medical and legal benchmarks, including a challenging multi-domain setting, show that DAMPER improves utility and rewrite quality over strong baselines while reducing privacy leakage.  
\section*{Acknowledgements}
This work is partially supported by Beijing Advanced Innovation Center for Future Blockchain and Privacy Computing, the National Natural Science Foundation of China  (U25B2070, 62372493), the Beijing Natural Science Foundation (Z230001).

\section*{Limitations} 
\vspace{-1mm}
As an early step toward automatic privacy span localization for client-side rewriting, our work has several limitations. We evaluate only two domains and a constructed multi-domain setting, and future work should extend to a broader range of domains. Moreover, our current setting focuses on a multi-domain query stream, where each input is assumed to admit an identifiable dominant domain; a more challenging direction is single-query multi-domain fusion, in which cues from different domains are intertwined within the same input and may interfere with each other. We plan to investigate mechanisms to disentangle such fine-grained interactions in our future work. 

\vspace{-1mm}
\section*{Ethical Considerations}
\vspace{-1mm}
This research aligns strictly with the ACL Ethics Policy. Our experiments rely exclusively on Pri-DDXPlus, Pri-SLJA, and Pri-Mixture, which are derived from established open-source repositories and strictly follow the data protocols set by prior work~\citep{zeng-etal-2025-privacyrestore}. We confirm that these benchmarks are fully anonymized and contain no real-world personally identifiable information (PII). Consequently, our study involves no interaction with human subjects, nor does it attempt to de-anonymize or reconstruct private data. It is important to note that our evaluation employs large language models (e.g., Qwen, GPT-4o). While these models facilitate automated rewriting and scoring, their outputs may reflect inherent biases or hallucinations common to pre-trained architectures.

\bibliography{custom}

\clearpage
\appendix
\appendixpage

\setcounter{tocdepth}{-1}
\renewcommand{\contentsname}{}
\addtocontents{toc}{\protect\setcounter{tocdepth}{2}}
\tableofcontents

\section{Notations} \label{app:notations}
Table~\ref{tab:notations} summarizes the notation used throughout this paper.

\begin{table*}[t]
\centering 
\fontsize{8}{8.5}\selectfont
\setlength{\tabcolsep}{5pt}  
\renewcommand{\arraystretch}{1.3} 
\begin{tabularx}{\linewidth}{l||X}
\noalign{\hrule height 1pt}
\rowcolor{gray!30}
\textbf{Symbol} & \textbf{Description} \\
\hline\hline
$\{a_i\}_{i=1}^M,\,M$ & Spans generated by TextChunker$(\cdot)$ for input text $x$ and their count \\
$\mathcal{D},\,\mathcal{D}_{\text{pref}}$ & Training corpus and preference dataset \\
$\mathcal{K},\,k$ & Total number of domains; domain index \\
$d_{k,i}$ & Affinity score between span $a_i$ and domain $k$, i.e., $d_{k,i}=\max_{1\le j\le J_k}\cos(\mathbf{z}_i,\mathbf{p}_{k,j})$ \\
$\hat{k}$ & Inferred domain index for input $x$ \\
$score_i$ & Privacy score for span $a_i$ (used for thresholding) \\
$\gamma_{k}$ & Privacy threshold for domain $k$ \\
$\hat{\mathcal{S}}_{\hat{k}}^x$ & Set of detected privacy spans in input $x$ under inferred domain $\hat{k}$ \\
$\mathcal{S}_k=\{s_{k,i}\}_{i=1}^{N_k},\,N_k$ & Annotated privacy spans and their count for domain $k$ (training set) \\
$\mathcal{S}_k^x=\{s_{k,i}^x\}_{i=1}^{N_k^x},\,N_k^x$ & Annotated privacy spans in input $x$ assigned to domain $k$ and their count \\
$\mathcal{S}_k^y=\{s_{k,i}^y\}_{i=1}^{N_k^x}$ & Corresponding set of rewritten spans aligned to $\mathcal{S}_k^x$ \\
$\mathcal{U}$ & Semantically meaningful spans \\
$h(\cdot),\,g(\cdot)$ & Trainable span encoder and frozen backbone encoder \\
$\mathbf{z}_{k,i}$ & Representation of $s_{k,i}$ encoded by $h(\cdot)$, i.e., $\mathbf{z}_{k,i}=h(s_{k,i})$ \\
$\mathbf{z}_{k,i}^x,\,\mathbf{z}_{k,i}^y$ & Representations of $s_{k,i}^x$ and $s_{k,i}^y$ encoded by $h(\cdot)$ \\
$\mathbf{z}_i$ & Representation of span $a_i$ encoded by $h(\cdot)$, i.e., $\mathbf{z}_i=h(a_i)$ \\
$\mathcal{A}_{k,i}$ & Positive set for anchor $s_{k,i}$ (in-domain privacy spans) \\
$\mathcal{N}_{k,i}$ & Negative set for anchor $s_{k,i}$ (out-of-domain privacy spans and non-privacy spans) \\
$\mathcal{G}_{k,i}$ & Union set $\mathcal{G}_{k,i}=\mathcal{A}_{k,i}\cup\mathcal{N}_{k,i}$ used in InfoNCE denominator \\
$\mathcal{L}_{CTR};\,\mathcal{L}_{DPO}$ & Multi-positive InfoNCE loss and DPO loss \\
$\tau_1$ & Temperature for contrastive loss \\
$\mathcal{P}_k=\{\mathbf{p}_{k,j}\}_{j=1}^{J_k},\,J_k$ & Set of privacy prototypes and the number of prototypes for domain $k$ \\
$j_{i}^{\star}$ & Index of the nearest prototype for the $i$-th rewritten span, $j_i^\star=\arg\max_{1\le j\le J_k}\cos(\mathbf{z}_{k,i}^y,\mathbf{p}_{k,j})$ \\
$\mathcal{Y}(x)=\{y^{(c)}\}_{c=1}^C$ & Set of candidate rewrites for input $x$ \\
$y_w,\,y_l$ & Preferred rewrite (winner) and dispreferred rewrite (loser) \\
$r_{\text{priv}}(\cdot),\,r_{\text{util}}(\cdot),\,r(\cdot)$ & Privacy reward, utility reward, and composite reward \\
\hline
$\alpha$ & Weighting parameter balancing privacy and utility rewards \\
$\beta$ & Hyperparameter for DPO \\
$\pi_{\theta},\,\pi_{ref}$ & Rewriter policy and reference policy \\
$x;\,y$ & Input text and rewritten text \\
$\mathcal{V}$ & Vocabulary of the rewriter $\pi_{\theta}$ \\
$u_{\ell},\,\bar u_{\ell}$ & Logit vector at decoding step $\ell$ and its clipped version, $\bar u_{\ell}=\mathrm{clip}(u_{\ell};R_1,R_2)$ \\
$R_1,\,R_2$ & Lower bound and upper bound for logit clipping \\
$\Delta u$ & Utility sensitivity \\
$\tau_2$ & Sampling temperature for the rewriter (DP mechanism) \\
$\epsilon_{token},\,\epsilon_{text},\,\epsilon$ & Token-level privacy cost and text-level (query-level) privacy budget, privacy hyperparameter \\
$n_{\text{sp}}(x),\,n_{\text{sp}}^{\max}$ & Number of regenerated tokens within detected spans in $x$; global upper bound used for budgeting \\
$\sigma(\cdot)$ & Sigmoid function \\
\noalign{\hrule height 0.8pt} 
\end{tabularx}
\caption{Summary of notation used in this paper.} 
\label{tab:notations}
\end{table*}

\section{Related Work}
\label{app:related_work}
\subsection{Token-level LDP Substitution}
Local Differential Privacy (LDP)~\citep{duchi2013local} has been widely used to privatize text at the token level.
Early mechanisms perturb each word independently and sample substitutes under metric-LDP constraints.
SanText~\citep{yue-etal-2021-differential} selects replacements according to embedding distance to provide provable token-level guarantees, while CusText~\citep{chen-etal-2023-customized} restricts the candidate neighborhood to reduce semantic drift.
A key limitation is that token-level substitution effectively treats every token as sensitive, often leading to indiscriminate perturbation and stylistic or pragmatic mismatch.

\subsection{Sequence-level DP Paraphrasing}
To better exploit context, sequence-level approaches rewrite entire sentences or documents with DP mechanisms.
DP-Paraphrase~\citep{mattern-etal-2022-limits} analyzes the limitations of token-level DP, highlighting unfavorable privacy--utility scaling with text length.
DP-BART~\citep{igamberdiev-habernal-2023-dp} improves fluency with an encoder--decoder architecture and training-time DP techniques.
DP-Prompt~\citep{utpala-etal-2023-locally} leverages LLM prompting for private rewriting without fine-tuning, and DP-MLM~\citep{meisenbacher-etal-2024-dp} proposes the 1-Diffractor mechanism to generate multiple DP-compliant candidates and select high-quality realizations.
Recent work explores strategies that combine multiple granularities.
DP-GTR~\citep{li-dpgtr-2025}, for example, generates multiple DP paraphrases and identifies consensus keywords that consistently survive privatization for iterative suppression.
Although these methods typically yield better coherence than token-level schemes, they still adopt a largely uniform notion of privacy across the input, hindering the tailoring of perturbations to regions with varying sensitivity.

\subsection{Span-level Privacy Rewriting} Span-level methods focus on localizing and sanitizing sensitive segments while preserving the integrity of the surrounding non-sensitive context. Conventional approaches typically couple Named Entity Recognition (NER) or PII classifiers with selective sanitization, applying DP mechanisms strictly to tokens flagged as sensitive~\citep{yue-etal-2021-differential,mattern-etal-2022-limits}. More advanced frameworks, such as PrivacyRestore~\citep{zeng-etal-2025-privacyrestore}, decompose the process into client-side removal and server-side restoration to enhance downstream utility. 
Recently, NaPaRe~\citep{huang2025zeroshot} introduced a zero-shot framework that utilizes iterative tree search for controllable rewriting (e.g., abstraction) without task-specific fine-tuning.
Similarly, DP-Fusion~\citep{thareja2025dp} proposes a differentially private inference mechanism that mixes output distributions from original and redacted contexts to provably bound the influence of identified sensitive token groups.

Despite these advancements in locality and controllability, existing pipelines face critical barriers to deployable client-side use. A substantial limitation lies in the rigidity of detection mechanisms. These methods inherently assume that sensitive spans can be exhaustively identified via static inventories or explicit user cues—which precludes autonomous localization and severely restricts adaptability to domain-contingent semantics. Beyond detection, distinct theoretical and architectural constraints persist. Recent search-based systems often prioritize empirical robustness against reconstruction attacks over rigorous formalism, lacking the formal span-level DP guarantees necessary for worst-case protection. Furthermore, frameworks like PrivacyRestore necessitate restoration mechanisms that are tightly coupled to specific server-side LLMs, preventing their deployment in model-agnostic scenarios where the downstream service is an opaque black-box API.

In contrast, our framework integrates domain-specific representation learning with prototype-guided localization and a DP-compliant rewriting mechanism. This design achieves autonomous, domain-contingent sanitization with formal privacy guarantees, while remaining fully decoupled from downstream LLM services.

\section{Preliminaries}
\label{sec:prelim}
\subsection{Differential Privacy}
Differential Privacy (DP)~\citep{dwork2006calibrating,dwork2014algorithmic} protects the contribution of any single individual in a dataset.
Let $\mathcal{X}$ denote the input space and $\mathcal{Y}$ the output space.
Two inputs $x, x' \in \mathcal{X}$ are called \emph{neighbors}, written $x \sim x'$, if they differ in the contribution of one individual (or, in the local setting, correspond to two possible private values of a single user).
A randomized mechanism $\mathcal{M}: \mathcal{X} \to \mathcal{Y}$ is said to satisfy $\epsilon$-DP if for all $x \sim x'$ and measurable $S \subseteq \mathcal{Y}$,
\begin{equation}
  \Pr[\mathcal{M}(x) \in S]
  \le e^\epsilon \Pr[\mathcal{M}(x') \in S].
\end{equation}
where $\epsilon > 0$ is the privacy budget.
In the context of text rewriting, we often operate under Local Differential Privacy (LDP)~\citep{duchi2013local}, where the sanitization mechanism is applied locally by the user before data leaves their device.

\subsection{The Exponential Mechanism}The Exponential Mechanism (EM)~\citep{mcsherry2007mechanism} is a fundamental technique for selecting outputs from a discrete domain $\mathcal{Y}$ while preserving differential privacy. Given an input $x$, a utility function $u: \mathcal{X} \times \mathcal{Y} \to \mathbb{R}$ assigns a real-valued score to each possible output $y \in \mathcal{Y}$, reflecting its quality or relevance. The privacy guarantee of the mechanism depends on the global sensitivity of the utility function, defined as the maximum change in utility caused by modifying a single input entry:
\begin{equation}
\label{eq:sensitivity}\Delta u = \max_{x \sim x'} \max_{y \in \mathcal{Y}} |u(x, y) - u(x', y)|.
\end{equation}
The Exponential Mechanism $\mathcal{M}_{\text{EM}}(x)$ selects an output $y$ with probability proportional to its utility, scaled by the privacy budget $\epsilon$ and sensitivity $\Delta u$:\begin{equation}\label{eq:em_prob}\Pr[\mathcal{M}_{\text{EM}}(x) = y] = \frac{\exp\left(\frac{\epsilon \cdot u(x, y)}{2\Delta u}\right)}{\sum_{y' \in \mathcal{Y}} \exp\left(\frac{\epsilon \cdot u(x, y')}{2\Delta u}\right)}.\end{equation}This distribution ensures that outputs with higher utility are exponentially more likely to be chosen, providing a theoretically grounded trade-off between utility and privacy.

\vspace{0.5mm} 
\noindent\textbf{Sampling-based Exponential Mechanism.}
In the context of autoregressive or masked language generation, the output domain corresponds to the vocabulary $\mathcal{V}$ at each decoding step. Recent works operationalize the Exponential Mechanism directly via the model's standard sampling process. Specifically, the unnormalized log-probability (logit) $v_t \in \mathbb{R}$ for a token $t \in \mathcal{V}$ serves as the utility function, i.e., $u(x, t) = v_t$. The standard softmax sampling distribution with a temperature $\tau > 0$ is given by:
\begin{equation}
\Pr(t|x) = \frac{\exp(v_t / \tau)}{\sum_{t' \in \mathcal{V}} \exp(v_{t'} / \tau)}.
\end{equation}
By equating the exponents in Eq.~\eqref{eq:em_prob} and the softmax function, we observe that setting the temperature $\tau = \frac{2\Delta u}{\epsilon}$ renders the sampling process formally equivalent to the Exponential Mechanism~\citep{mattern-etal-2022-limits, utpala-etal-2023-locally}. However, the raw logits of neural language models are theoretically unbounded, implying an infinite sensitivity $\Delta u$. To satisfy DP guarantees, it is necessary to bound the utility function. This is typically achieved by clipping the logits to a fixed range $[R_1, R_2]$ (or applying a norm bound) before sampling. Under this constraint, the sensitivity is upper-bounded by $\Delta u \le 2(R_2-R_1)$ (the maximum distance between any two values in $[R_1, R_2]$). Thus, sampling from the clipped logits with temperature $\tau$ satisfies $\epsilon$-DP at the token level, where:
\begin{equation}
\epsilon = \frac{4(R_2-R_1)}{\tau}.
\end{equation}
This formulation allows standard LLMs to function as differentially private generators by simply adjusting the sampling temperature and enforcing logit constraints.

\subsection{Prototype Learning}
Prototype learning maps inputs into a latent embedding space where classes or clusters are represented by identifying \emph{prototypes}. In supervised settings, such as Prototypical Networks~\citep{snell2017prototypical}, prototypes are typically computed as the mean embedding of support examples for each class. In unsupervised or semi-supervised settings, prototypes can be derived via clustering algorithms (e.g., k-means~\citep{mcqueen1967some}, GMM~\citep{dempster1977maximum} or FINCH~\citep{sarfraz2019efficient}) to serve as structured, discrete summaries of the underlying data distribution. 

\subsection{Direct Preference Optimization}
\label{sec:prelim-dpo} 
Preference alignment aims to steer language models toward desired behaviors using relative feedback rather than absolute rewards~\citep{christiano2017deep,ouyang2022training,bai2022training}. Direct Preference Optimization (DPO)~\citep{rafailov2023direct} optimizes a policy $\pi_\theta$ directly from a dataset of preference pairs $(x, y_w, y_l)$, where $y_w$ is preferred over $y_l$.
Unlike Reinforcement Learning from Human Feedback (RLHF)~\citep{ouyang2022training}, which requires training a separate reward model, DPO derives a closed-form objective by implicitly defining the reward via the ratio of the policy likelihood to a frozen reference model $\pi_{\text{ref}}$. The objective minimizes the negative log-likelihood of the preference data:
\begin{equation}
\label{eq:dpo_loss} 
-\mathbb{E} \Big[ \log \sigma \Big( \beta \log \frac{\pi_\theta(y_w|x)}{\pi_{\text{ref}}(y_w|x)} - \beta \log \frac{\pi_\theta(y_l|x)}{\pi_{\text{ref}}(y_l|x)} \Big) \Big].
\end{equation}
where $\beta$ is a hyperparameter controlling the deviation from the reference policy. This approach provides a stable and computationally efficient method for aligning model outputs with complex preference criteria.

\section{Details of TextChunker}
\label{app:TextChunker}
We adopt a rule-based and syntactically guided text chunking approach to decompose the input text into semantically coherent spans. Given an input text, we first perform sentence segmentation and shallow syntactic analysis using spaCy with the \texttt{en\_core\_web\_sm} model. These annotations provide sufficient structural information for phrase-level reasoning without requiring full semantic parsing. Inspired by~\citet{lin2020joint}, for each sentence, we then identify a small set of semantic frame triggers, implemented as regular-expression patterns. Once a trigger is detected, the text following the trigger is treated as a candidate enumeration region and is split using weak separators, including commas, semicolons, and coordinating conjunctions (\emph{and}, \emph{or}). This design enables robust handling of parallel structures while avoiding overly aggressive segmentation. To improve recall under more complex or implicit sentence constructions, we additionally incorporate syntax-aware phrase extraction. Concretely, we collect noun phrases identified by spaCy’s noun chunker, gerundive verb phrases whose head tokens are tagged as VBG~\citep{marcus1993building} (with the span defined by the corresponding dependency subtree), and infinitive purpose constructions matching the pattern \emph{to verb}. These phrase types are widely recognized as minimal semantic carriers in span-based modeling and information extraction~\citep{joshi2020spanbert,chen-etal-2023-frustratingly}. Finally, we prevent spans from crossing sentence-internal strong punctuation and normalize span boundaries by removing leading conjunctions and trailing punctuation.

\section{Algorithms}
\label{app:algorithm}
The algorithmic details of DAMPER are presented below. Algorithm~\ref{alg:training_master} outlines the Offline Training Phase, coordinating prototype learning (Algorithm~\ref{alg:prototypes}) and preference construction (Algorithm~\ref{alg:preference}). The Online Inference Phase is detailed in Algorithm~\ref{alg:inference}.

\begin{algorithm}[t]
\caption{Learning Domain Privacy Prototypes}
\label{alg:prototypes}
\begin{algorithmic}[1]
\REQUIRE Annotated privacy spans \(\{\mathcal{S}_k\}\), backbone \(g(\cdot)\), temperature \(\tau_1\).
\ENSURE Privacy prototypes \(\{\mathcal{P}_k\}_{k \in \mathcal{K}}\)
\STATE Initialize span encoder \(h(\cdot)\) from \(g(\cdot)\).
\STATE // Contrastive Representation Learning
\FOR{each training step}
    \STATE Sample batch of anchor spans \(s_{k,i}\);
    \STATE Get positive set \(\mathcal{A}_{k,i}\) and negative set \(\mathcal{G}_{k,i}\);
    \STATE Update \(h(\cdot)\) by minimizing \(\mathcal{L}_{\mathrm{CTR}}\) (Eq.~\eqref{eq:infonceloss}).
\ENDFOR
\STATE // Prototype Clustering
\FOR{each domain \(k \in \mathcal{K}\)}
    \STATE Extract \(\mathbf{z}_k = \{h(s) \mid s \in \mathcal{S}_k\}\).
    \STATE Cluster \(\mathcal{P}_k \leftarrow \textsc{Finch}(\mathbf{z}_k)\).
\ENDFOR
\RETURN \(\{\mathcal{P}_k\}\)
\end{algorithmic}
\end{algorithm}

\begin{algorithm}[t]
\caption{Preference Dataset Construction}
\label{alg:preference}
\begin{algorithmic}[1]
\REQUIRE Training corpus \(\mathcal{D}\), reference model \(\pi_{\mathrm{ref}}\), prototypes \(\{\mathcal{P}_k\}\), span encoder \(h(\cdot)\), hyperparameter \(\alpha\).
\ENSURE Preference dataset \(\mathcal{D}_{\mathrm{pref}}\).

\STATE Initialize \(\mathcal{D}_{\mathrm{pref}} \leftarrow \emptyset\).
\FOR{each input \(x\) in \(\mathcal{D}\)}
    \STATE \(\mathcal{Y}(x) \leftarrow \{y^{(n)} \sim \pi_{\mathrm{ref}}(x, \mathcal{S}_k^x)\}_{n=1}^N\).
    \STATE // Composite Reward Scoring
    \FOR{each candidate \(y \in \mathcal{Y}(x)\)}
        \STATE Compute \(r_{\mathrm{priv}}(y)\) via Eq.~\eqref{eq:reward_priv};
        \STATE Compute \(r_{\mathrm{util}}(y)\) via Eq.~\eqref{eq:reward_util};
        \STATE \(r(y) \leftarrow (1-\alpha)r_{\mathrm{priv}}(y) + \alpha r_{\mathrm{util}}(y)\).
    \ENDFOR
    \STATE Select \(y_w \leftarrow \arg\max_{y \in \mathcal{Y}(x)} r(y)\);
    \STATE Select \(y_l \leftarrow \arg\min_{y \in \mathcal{Y}(x)} r(y)\);
    \STATE \(\mathcal{D}_{\mathrm{pref}} \leftarrow \mathcal{D}_{\mathrm{pref}} \cup \{(x, y_w, y_l)\}\).
\ENDFOR
\RETURN \(\mathcal{D}_{\mathrm{pref}}\)
\end{algorithmic}
\end{algorithm}

\begin{algorithm}[t]
\caption{DAMPER Offline Training Phase}
\label{alg:training_master}
\begin{algorithmic}[1]
\REQUIRE Training corpus \(\mathcal{D}\), annotated spans \(\{\mathcal{S}_k\}\), backbone \(g(\cdot)\), reference model \(\pi_{\mathrm{ref}}\).
\ENSURE Optimized rewriter \(\pi_{\theta}\).

\STATE \textbf{// Prototype Learning}
\STATE \(\{\mathcal{P}_k\}, h(\cdot) \leftarrow \text{Algorithm~\ref{alg:prototypes}}(\{\mathcal{S}_k\}, g)\)

\STATE \textbf{// Preference Construction}
\STATE \(\mathcal{D}_{\mathrm{pref}} \leftarrow     \text{Algorithm}~\ref{alg:preference}(\mathcal{D}, \pi_{\mathrm{ref}}, \{\mathcal{P}_k\}, h)\)

\STATE \textbf{// Preference Alignment}
\STATE Initialize \(\pi_{\theta}\) from \(\pi_{\mathrm{ref}}\).
\WHILE{not converged}
    \STATE Sample batch \((x, y_w, y_l)\) from \(\mathcal{D}_{\mathrm{pref}}\).
    \STATE Update \(\pi_{\theta}\) by minimizing \(\mathcal{L}_{\mathrm{DPO}}\) (Eq.~\eqref{eq:dpoloss}).
\ENDWHILE

\RETURN \(\pi_{\theta}\)
\end{algorithmic}
\end{algorithm}
 
\begin{algorithm}[t]
\caption{Online Inference Phase}
\label{alg:inference}
\begin{algorithmic}[1]
\REQUIRE User query \(x\), span encoder \(h(\cdot)\), prototypes \(\{\mathcal{P}_k\}\), rewriter \(\pi_{\theta}\).
\ENSURE rewritten query \(y\).

\STATE \textbf{// Text Segmentation}
\STATE Segment \(\{a_i\}_{i=1}^M \leftarrow \textsc{TextChunker}(x)\);
\STATE Compute \(\mathbf{z}_i \leftarrow h(a_i)\) for \(i=1 \dots M\).

\STATE \textbf{// Domain Inference} 
\FOR{each domain \(k \in \mathcal{K}\)}
    \FOR{each span \(i = 1 \dots M\)}
        \STATE Calculate \(d_{k,i} \leftarrow \max_{\mathbf{p} \in \mathcal{P}_k} \cos(\mathbf{z}_i, \mathbf{p})\).
    \ENDFOR
    \STATE Domain score: \(s_k \leftarrow \frac{1}{M} \sum_{i=1}^M d_{k,i}\).
\ENDFOR
\STATE Infer global domain: \(\hat{k} \leftarrow \arg\max_{k \in \mathcal{K}} s_k\).

\STATE \textbf{// Privacy Span Localization} 
\STATE Initialize privacy set \(\hat{\mathcal{S}}^x \leftarrow \emptyset\).
\STATE sort affinities: \(d_{\hat{k},(1)} \leq d_{\hat{k},(1)} \leq \cdots \leq d_{\hat{k},(M)}\)
\FOR{\(t=1 \dots M\)}
    \STATE Calculate \(\sigma^2(t)\) by Eq.~\eqref{eq:sigmat}.
\ENDFOR
\STATE Find \(t^\star \leftarrow \arg\max_t \sigma^2(t)\).
\STATE Threshold: \(\gamma_{\hat{k}} \leftarrow (d_{\hat{k},(t^\star)}+d_{\hat{k},(t^{\star+1})})/2\)
\FOR{each span \(i = 1 \dots M\)}
    \IF{\(d_{\hat{k},i} > \gamma_{\hat{k}}\)}
        \STATE \(\hat{\mathcal{S}}^x \leftarrow \hat{\mathcal{S}}^x \cup \{a_i\}\).
    \ENDIF
\ENDFOR

\STATE \textbf{// DP Span-level Rewriting}
\STATE Apply rewriter only on detected spans:
\STATE \(\tilde{y} = \tilde{\pi}_\theta(x, \hat{\mathcal{S}}_{\hat{k}}^x)\) with EM (Eq.~\eqref{eq:softmax-sampling-rewrite}).
\end{algorithmic}
\end{algorithm}
 
\section{More Experimental Setups}\label{app:expsetup}

\subsection{More Details of Datasets}\label{app:datasets}
We evaluate our framework on two established domain-specific datasets and a synthesized multi-domain benchmark constructed as follows.

\vspace{0.5mm}
\noindent\textbf{Pri-DDXPlus.}
This dataset focuses on the medical diagnosis domain~\citep{zeng-etal-2025-privacyrestore}. Each sample consists of a patient symptom description, explicitly annotated with privacy-sensitive and non-sensitive spans based on medical confidentiality standards. The task is formulated as a multiple-choice question, where each sample is associated with one correct diagnosis and three randomly sampled incorrect diagnoses as in-domain distractors.

\vspace{0.5mm}
\noindent\textbf{Pri-SLJA.}
Analogous to Pri-DDXPlus, this dataset targets the legal judgment domain~\citep{zeng-etal-2025-privacyrestore}. Each entry contains a detailed description of a legal case, with span-level annotations distinguishing privacy-sensitive details from general context according to legal norms. Similarly, each sample is paired with one correct judgment and three random incorrect judgments.

\vspace{0.5mm}
\noindent\textbf{Pri-Mixture.}
To assess performance in multi-domain scenarios, we construct Pri-Mixture by augmenting the two base datasets with multi-domain distractors. Specifically, we append four law-related options to the original medical samples (Pri-DDXPlus) and four medical-related options to the legal samples (Pri-SLJA). These augmented datasets are then amalgamated to form a unified multi-domain benchmark, challenging the model to distinguish correct answers amidst both in-domain and out-of-domain noise.

\subsection{More Details of Metrics}\label{app:metrics}
We employ a comprehensive suite of metrics to evaluate our framework across three critical dimensions: downstream utility, semantic fidelity, and privacy preservation.

\vspace{0.5mm}
\noindent\textbf{Accuracy (Acc \(\uparrow\)).}
To assess the utility of the rewritten text for downstream applications, we measure the prediction accuracy of a cloud-based LLM on the processed queries. This evaluation is standardized using a structured prompt template (see Appendix~\ref{app:template_acc}) to ensure consistent decision-making logic across different test cases.

\vspace{0.5mm}
\noindent\textbf{BERTScore (BS \(\uparrow\)).}
BERTScore quantifies the semantic similarity between the original query $x$ and the rewritten output $y$. By computing the cosine similarity of their contextual embeddings via a pre-trained encoder, it serves as a robust proxy for measuring how well the semantic content is preserved after rewriting.

\vspace{0.5mm}
\noindent\textbf{LLM-Judge (LLM-J \(\uparrow\)).}
Following recent benchmarks~\citep{zheng2023judging, zeng-etal-2025-privacyrestore}, we employ an LLM-Judge to evaluate the overall quality and coherence of the generated text. Specifically, we first generate responses using Qwen2.5-7B-Instruct~\citep{qwen2.5,qwen2} based on the rewritten queries, and then leverage DeepSeek-V3~\citep{deepseekai2025deepseekv32} as an evaluator to assign a quality score on a scale of 1 to 10. The prompt templates used for this evaluation are provided in Appendix~\ref{app:template_inf_rate}.

\vspace{0.5mm}
\noindent\textbf{Semantic Obfuscation Index (SOI \(\uparrow\)) and Domain Fidelity Score (DFS \(\uparrow\)).}
These dual metrics quantify the effectiveness of the rewriting policy in balancing competing objectives. \textbf{SOI} measures the degree of privacy masking, directly corresponding to the privacy reward defined in Eq.~\eqref{eq:reward_priv}, while \textbf{DFS} assesses the preservation of domain-specific characteristics, aligning with the utility reward in Eq.~\eqref{eq:reward_util}. To ensure a focused assessment, both metrics are computed exclusively on the localized privacy spans rather than the entire text.

\vspace{0.5mm}
\noindent\textbf{SOI-Drop (\(\downarrow\)) and DFS-Drop (\(\downarrow\)).}
To analyze the sensitivity of model performance to the trade-off parameter $\alpha$, we introduce the Drop metrics. Let $\mathrm{M} \in \{\mathrm{SOI}, \mathrm{DFS}\}$ denote the min-max normalized score for a given metric. The performance drop is defined as the deviation from the optimal achievable value:
\begin{equation}\label{eq:drop}
\mathrm{M\text{-}Drop} = \frac{\max(\mathrm{M}) - \mathrm{M}}{\max(\mathrm{M})}.
\end{equation}
This metric helps in identifying the parameter configuration that minimizes the collective loss in privacy and utility.

\vspace{0.5mm}
\noindent\textbf{Privacy F1-Score (PF1 $\uparrow$).}
To provide a holistic assessment of the span localization module, we employ the F1-Score, which balances the trade-off between Precision and Recall. Precision quantifies the proportion of predicted spans that are actual ground-truth privacy spans:
\begin{equation}
\mathrm{Precision} = \frac{|S_k^x \cap \hat{S}_{\hat{k}}^x|}{|\hat{S}_{\hat{k}}^x|},
\end{equation}
while Recall measures the proportion of ground-truth spans that are successfully retrieved by the model:
\begin{equation}
\mathrm{Recall} = \frac{|S_k^x \cap \hat{S}_{\hat{k}}^x|}{|S_k^x|}.
\end{equation} 
The unified PF1 metric is calculated as the harmonic mean:
\begin{equation}
\label{eq:pf1}
\mathrm{PF1} = 2 \cdot \frac{\mathrm{Precision} \cdot \mathrm{Recall}}{\mathrm{Precision} + \mathrm{Recall}}.
\end{equation}

\vspace{0.5mm}
\noindent\textbf{ROUGE-L ($\downarrow$).}
ROUGE-L~\citep{Lin2004ROUGEAP} measures the structural similarity between the generated sequence and the reference sequence based on the longest common subsequence. In our privacy context, we calculate this metric between the original sensitive spans and their rewritten counterparts. A lower ROUGE-L score indicates less lexical overlap, reflecting more effective obfuscation of sensitive information.

\subsection{More Details of Implementation}
\label{app:implementationdetails}
\noindent\textbf{DAMPER Training Configurations.}
The span encoder is fine-tuned on 149 medical and 142 legal privacy span types. We apply LoRA with a rank of 16 and train for 30 epochs using a batch size of 32 and the AdamW optimizer (learning rate $10^{-4}$). For preference construction and DPO, we utilize LLaMA-Factory~\citep{zheng2024llamafactory} to generate 10 candidates per sample. The policy model is fine-tuned for 7 epochs with LoRA (rank 16), a batch size of 16, and a learning rate of $5 \times 10^{-6}$. Regarding dataset-specific hyperparameters, we set $R_1=5$ and $R_2=20$.
 
\vspace{0.5mm}
\noindent\textbf{Baseline Configurations.}
To ensure a fair comparison, all baselines are evaluated under the same total privacy budget constraint ($\epsilon_\text{text}$) as DAMPER. For \textbf{DP-Paraphrase}, we adopt GPT-2, fine-tuned by \citet{meisenbacher-etal-2024-dp}, as the paraphrasing model. For \textbf{DP-Prompt}, we utilize the open-source Flan-T5-xl~\citep{https://doi.org/10.48550/arxiv.2210.11416} to balance performance and adaptability. Regarding \textbf{DP-MLM} and \textbf{DP-MLM$_{\text{oracle}}$}, we employ RoBERTa-base~\citep{Liu2019RoBERTaAR}, concatenating the original text as auxiliary guidance during inference. Finally, for \textbf{PrivacyRestore}, to maintain comparability with DAMPER, we use Qwen2.5-7B-Instruct as the inference model and BERT-base-uncased~\citep{devlin2019bert} as the auxiliary retrieval model, with the number of common attention heads set to 175. All remaining configurations also follow the default settings reported in their original papers.

\begin{table*}[t] 
\centering
\small
\renewcommand{\arraystretch}{1.25}
\fontsize{8}{8}\selectfont
\setlength{\tabcolsep}{1pt}
\begin{tabularx}{\linewidth}{
l||*{2}{>{\centering\arraybackslash}X}|
   *{2}{>{\centering\arraybackslash}X}|
   *{2}{>{\centering\arraybackslash}X}|
   *{1}{>{\centering\arraybackslash}X}}
\noalign{\hrule height 1pt}
\rowcolor{gray!25} 
& \multicolumn{2}{c|}{\textbf{Full-Text Level}} & \multicolumn{2}{c|}{\textbf{Span Level (oracle)}} & \multicolumn{2}{c|}{\textbf{Span Level (auto)}} & \\
\cline{2-7} 
\rowcolor{gray!25}
\multicolumn{1}{c||}{\multirow{-2}{*}{Datasets}} & $n^{\text{max}}$ & $\epsilon_{\text{token}}$ & $n_{\text{sp}}^{\text{max}}$ & $\epsilon_{\text{token}}$ &  $n_{\text{sp}}^{\text{max}}$ & $\epsilon_{\text{token}}$ & \multicolumn{1}{c}{\multirow{-2}{*}{$\epsilon_{\text{text}}$}} \\
\hline
Pri-DDXPlus & 86 & 1.74 & 49 & 3.06 & 52 & 2.88 & 150 \\
Pri-SLJA & 178 & 0.84 & 41 & 3.66 & 104 & 1.44 & 150 \\
\noalign{\hrule height 1pt}
\end{tabularx}
\caption{Privacy budget settings of different baselines across Pri-DDXPlus and Pri-SLJA. Baselines are categorized into three levels: \textbf{Full-Text Level}, including DP-Paraphrase, DP-Prompt, and DP-MLM; \textbf{Span Level (oracle)}, including DP-MLM$_\text{oracle}$ and DAMPER$_\text{oracle}$; and \textbf{Span Level (auto)}, including DP-MLM$_\text{auto}$ and DAMPER.}
\label{tab:epsilonanalysis}
\end{table*}

\vspace{0.5mm}
\noindent\textbf{Privacy Budget Allocation.}
Following PrivacyRestore~\citep{zeng-etal-2025-privacyrestore}, we fix privacy hyperparameter $\epsilon$ to 75 and report $\epsilon_\text{text}$ and $\epsilon_{\text{token}}$ under $\epsilon$. Table~\ref{tab:epsilonanalysis} provides a detailed breakdown of the privacy budget allocation under the setting $\epsilon_{\text{text}}=2\epsilon$. Since baselines operate at different granularities (Full-text vs. Span-level), we convert the total budget $\epsilon_\text{text}$ into equivalent per-token budgets $\epsilon_{\text{token}}$ based on the maximum sequence lengths $n^{\text{max}}$ or maximum span lengths $n_{\text{sp}}^{\text{max}}$. 

All experiments were conducted under Python 3.12 on a system equipped with two \textbf{NVIDIA RTX A6000 GPUs} (48 GB each).

\section{Additional Experimental Results}\label{app:exps-results}

\subsection{More results of Performance Comparsion}\label{app:performance}
\begin{table*}[t] 
\centering
\small
\renewcommand{\arraystretch}{1.25}
\fontsize{8}{7.8}\selectfont
\setlength{\tabcolsep}{1pt}
\begin{tabularx}{\linewidth}{
l||*{3}{>{\centering\arraybackslash}X}|
   *{3}{>{\centering\arraybackslash}X}|
   *{3}{>{\centering\arraybackslash}X}}
\noalign{\hrule height 1pt}
\rowcolor{gray!25}
& \multicolumn{3}{c|}{\textbf{Pri-DDXPlus}} & \multicolumn{3}{c|}{\textbf{Pri-SLJA}} & \multicolumn{3}{c}{\textbf{Pri-Mixture}} \\
\cline{2-4} \cline{5-7} \cline{8-10} 
\rowcolor{gray!25}
\multicolumn{1}{c||}{\multirow{-2}{*}{Methods}} & ACC ${\uparrow}$ & BS ${\uparrow}$ & LLM-J ${\uparrow}$ & ACC ${\uparrow}$ & BS ${\uparrow}$ & LLM-J ${\uparrow}$ & ACC ${\uparrow}$ & BS ${\uparrow}$ & LLM-J ${\uparrow}$ \\
\hline\hline
No Rewriting & 87.60$_{\pm 0.07}$ & 1.00$_{\pm 0.00}$ & 5.60$_{\pm 0.14}$ & 89.79$_{\pm 0.28}$ & 1.00$_{\pm 0.00}$ & 6.89$_{\pm 0.08}$ & 88.45$_{\pm 0.14}$ & 1.00$_{\pm 0.00}$ & 6.04$_{\pm 0.12}$ \\
\hline\hline
\rowcolor{gray!10}
DP-Paraphrase & 46.20$_{\pm 0.08}$ & 0.44$_{\pm 0.00}$ & 1.74$_{\pm 0.01}$ & 47.45$_{\pm 0.13}$ & 0.41$_{\pm 0.00}$ & 1.95$_{\pm 0.07}$ & 43.49$_{\pm 0.19}$ & 0.43$_{\pm 0.00}$ & 1.81$_{\pm 0.02}$ \\
DP-Prompt & 30.37$_{\pm 1.29}$ & 0.29$_{\pm 0.00}$ & 1.72$_{\pm 0.08}$ & 25.15$_{\pm 0.83}$ & 0.23$_{\pm 0.00}$ & 2.24$_{\pm 0.13}$ & 27.16$_{\pm 1.09}$ & 0.27$_{\pm 0.00}$ & 1.90$_{\pm 0.01}$ \\
\rowcolor{gray!10}
DP-MLM & 34.40$_{\pm 0.96}$ & 0.40$_{\pm 0.03}$ & 1.86$_{\pm 0.04}$ & 29.81$_{\pm 0.42}$ & 0.29$_{\pm 0.08}$ & 2.41$_{\pm 0.16}$ & 31.16$_{\pm 0.28}$ & 0.35$_{\pm 0.04}$ & 2.05$_{\pm 0.02}$ \\
DP-MLM$_{\text{oracle}}$ & 52.19$_{\pm 1.77}$ & \underline{0.60}$_{\pm 0.03}$ & 3.87$_{\pm 0.08}$ & \underline{81.14}$_{\pm 1.67}$ & \textbf{0.75}$_{\pm 0.01}$ & 5.07$_{\pm 0.16}$ & 61.27$_{\pm 1.44}$ & \underline{0.66}$_{\pm 0.01}$ & 4.28$_{\pm 0.00}$ \\
\rowcolor{gray!10}
PrivacyRestore & \underline{76.79}$_{\pm 0.85}$ & - & \underline{4.41}$_{\pm 0.13}$ & 79.36$_{\pm 0.54}$ & - & \textbf{5.62}$_{\pm 0.04}$ & \underline{77.32}$_{\pm 0.64}$ & - & \underline{4.82}$_{\pm 0.07}$  \\
\hline
\rowcolor{lightblue}
DAMPER & \textbf{79.71}$_{\pm 0.03}$ & \textbf{0.71}$_{\pm 0.00}$ & \textbf{4.96}$_{\pm 0.05}$ & \textbf{83.17}$_{\pm 0.78}$ & \underline{0.68}$_{\pm 0.00}$ & \underline{5.53}$_{\pm 0.02}$ & \textbf{80.01}$_{\pm 0.41}$ & \textbf{0.69}$_{\pm 0.00}$ & \textbf{5.15}$_{\pm 0.02}$ \\

\noalign{\hrule height 0.8pt} 
\end{tabularx}
\caption{More results of performance comparison on three datasets. ACC is reported in \%, where ``-'' denotes unavailable results. Results are averaged over 3 runs. The \textbf{best} and \underline{second} are marked.}
\label{tab:app_main_experiment}
\end{table*}

We further conduct additional experiments under different privacy budget configurations. Specifically, the $\epsilon_{\text{token}}$ of DAMPER is set to 9.62 on Pri-DDXPlus and 4.81 on Pri-SLJA. The setting of $\epsilon_{\text{text}}$ and the $\epsilon_{\text{token}}$ for the other baselines follow the same protocol as in Table~\ref{tab:epsilonanalysis}. The results are reported in Table~\ref{tab:app_main_experiment}.

\subsection{Single-Domain Training}\label{app:singledomain}
We use the single dataset during training phase. As shown in Table~\ref{tab:singledomain}, we compare the evaluation results of single-domain training with multi-domain training of our method and other baselines. The performance on the two single-domain datasets is nearly comparable to that achieved with multi-domain training. On Pri-DDXPlus, a slight degradation is observed in both rewriting quality and task accuracy, whereas on Pri-SLJA, single-domain training yields marginally better performance. These results indicate that DAMPER can effectively adapt to multi-domain rewriting tasks. 
\begin{table}[t]
\centering
\vspace{-2mm}
\setlength{\tabcolsep}{3pt}
\renewcommand{\arraystretch}{1.3}
\fontsize{8}{8}\selectfont
\begin{tabularx}{\columnwidth}{l||YYY}
\noalign{\hrule height 0.8pt} 
\rowcolor{gray!25} 
& \multicolumn{3}{c}{\textbf{Pri-DDXPlus}} \\ 
\cline{2-4}
\rowcolor{gray!25}
\multirow{-2}{*}{Methods} & ACC ${\uparrow}$ & BS ${\uparrow}$ & LLM-J ${\uparrow}$\\
\hline\hline
DP-MLM$_{\text{oracle}}$ & 50.94 & 0.59 & 3.61 \\
PrivacyRestore & 75.82 & - & 4.32 \\
\hline
DAMPER$_{\text{multi}}$ & \textbf{78.13} & \textbf{0.68} & \textbf{4.76}\\
DAMPER$_{\text{single}}$ & 77.83 & 0.61 & 4.69\\
\noalign{\hrule height 0.8pt} 
\rowcolor{gray!25} 
& \multicolumn{3}{c}{\textbf{Pri-SLJA}} \\ 
\cline{2-4}
\rowcolor{gray!25}
\multirow{-2}{*}{Methods} & ACC ${\uparrow}$ & BS ${\uparrow}$ & LLM-J ${\uparrow}$\\
\hline\hline
DP-MLM$_{\text{oracle}}$ & 78.38 & \textbf{0.73} & 4.81 \\
PrivacyRestore & 78.38 & - & 5.26 \\
\hline
DAMPER$_{\text{multi}}$ & 82.68 & 0.66 & 5.37\\
DAMPER$_{\text{single}}$ & \textbf{82.71} & 0.55 & \textbf{5.66}\\
\noalign{\hrule height 0.8pt} 
\end{tabularx}
\caption{Performance comparison on the single-domain. \textbf{DAMPER$_{\textbf{multi}}$} denotes the setting where both Pri-DDXPlus and Pri-SLJA are jointly used during training (evaluated on the corresponding dataset), while \textbf{DAMPER$_{\textbf{single}}$} refers to training using only a single dataset.}
\label{tab:singledomain}
\end{table}



\subsection{More Results of Ablation Study}\label{app:ablation_full}
In this section, we provide comprehensive ablation results for the individual datasets, \textbf{Pri-DDXPlus} and \textbf{Pri-SLJA}, which align with the mixture dataset observations presented in the main text (Table~\ref{tab:ablation_appendix}). On \textbf{Pri-DDXPlus}, the combination of $\mathcal{L}_{CTR}$ and $\mathcal{L}_{DPO}$ yields an ACC improvement of 44.02
\% over the baseline, surpassing the gains from using either module individually. Consistent with this trend, the combined approach on \textbf{Pri-SLJA} achieves the highest performance across all metrics, reaching an ACC of 82.68\% and a BS of 0.66. These results align with the analysis on Pri-Mixture, robustly validating the effectiveness and necessity of each component in our DAMPER.

\begin{table}[t]
\centering 
\setlength{\tabcolsep}{3pt}
\renewcommand{\arraystretch}{1.25}
\fontsize{8}{7.8}\selectfont
\begin{tabularx}{\columnwidth}{cc||YYY}
\noalign{\hrule height 0.8pt} 
\rowcolor{gray!25} 
& & \multicolumn{3}{c}{\textbf{Pri-DDXPlus}} \\ 
\cline{3-5}
\rowcolor{gray!25}
\multirow{-2}{*}{$\mathcal{L}_{CTR}$} & \multirow{-2}{*}{$\mathcal{L}_{DPO}$} & ACC ${\uparrow}$ & BS ${\uparrow}$ & LLM-J ${\uparrow}$\\
\hline\hline
                &                & 34.11 & 0.42 & 2.22 \\
                & \checkmark     & 66.50$_{\scriptscriptstyle \textcolor{blue}{\uparrow 32.39}}$
                                & 0.48$_{\scriptscriptstyle \textcolor{blue}{\uparrow 0.06}}$
                                & 3.83$_{\scriptscriptstyle \textcolor{blue}{\uparrow 1.61}}$ \\
\checkmark      &                & 65.59$_{\scriptscriptstyle \textcolor{blue}{\uparrow 31.48}}$
                                & 0.61$_{\scriptscriptstyle \textcolor{blue}{\uparrow 0.19}}$
                                & 3.91$_{\scriptscriptstyle \textcolor{blue}{\uparrow 1.69}}$ \\
\checkmark      & \checkmark     & \textbf{78.13}$_{\scriptscriptstyle \textcolor{blue}{\uparrow 44.02}}$
                                & \textbf{0.68}$_{\scriptscriptstyle \textcolor{blue}{\uparrow 0.26}}$
                                & \textbf{4.76}$_{\scriptscriptstyle \textcolor{blue}{\uparrow 2.54}}$ \\
\noalign{\hrule height 0.8pt} 
\rowcolor{gray!25} 
& & \multicolumn{3}{c}{\textbf{Pri-SLJA}} \\ 
\cline{3-5}
\rowcolor{gray!25}
\multirow{-2}{*}{$\mathcal{L}_{CTR}$} & \multirow{-2}{*}{$\mathcal{L}_{DPO}$} & ACC ${\uparrow}$ & BS ${\uparrow}$ & LLM-J ${\uparrow}$ \\
\hline\hline
                &                & 40.08 & 0.36 & 2.47 \\
                & \checkmark     & 74.46$_{\scriptscriptstyle \textcolor{blue}{\uparrow 34.38}}$
                                & 0.39$_{\scriptscriptstyle \textcolor{blue}{\uparrow 0.03}}$
                                & 4.94$_{\scriptscriptstyle \textcolor{blue}{\uparrow 2.47}}$ \\
\checkmark      &                & 69.49$_{\scriptscriptstyle \textcolor{blue}{\uparrow 29.41}}$
                                & 0.56$_{\scriptscriptstyle \textcolor{blue}{\uparrow 0.20}}$
                                & 4.53$_{\scriptscriptstyle \textcolor{blue}{\uparrow 2.06}}$ \\
\checkmark      & \checkmark     & \textbf{82.68}$_{\scriptscriptstyle \textcolor{blue}{\uparrow 42.60}}$
                                & \textbf{0.66}$_{\scriptscriptstyle \textcolor{blue}{\uparrow 0.30}}$
                                & \textbf{5.37}$_{\scriptscriptstyle \textcolor{blue}{\uparrow 2.90}}$ \\
\noalign{\hrule height 1pt}
\end{tabularx}
\caption{\textbf{Ablation study} on the performance of $\mathcal{L}_{CTR}$ and $\mathcal{L}_{DPO}$. The \textbf{best} are marked.}
\label{tab:ablation_appendix}
\end{table}

\subsection{Results of TextChunker's Robustness}\label{app:textchunkerrobustness}
We conducted a robustness study on segmentation by replacing the original TextChunker with alternative span proposal strategies. Specifically, we considered the following variants. \textbf{Boundary Perturbation (BP)}: Starting from the original TextChunker outputs, we randomly select a proportion $p$ (in our setting, $p = 0.2$) of spans and apply random re-segmentation. \textbf{YAKE-based Segmentation (YAKE)}\citep{campos2020yake}: We use YAKE, an unsupervised keyphrase extraction method that scores candidate phrases based on local statistical features such as term frequency, positional information, and contextual distribution. The extracted keyphrases are treated as candidate spans. \textbf{n-gram Enumeration (n-gram)}: We enumerate all contiguous n-grams with length $L \leq 6$ as candidate spans. This strategy increases coverage while introducing denser span proposals. The experimental results in Table~\ref{tab:textchunker} show that, on both Pri-DDXPlus and Pri-SLJA, DAMPER achieves comparable ACC and PF1 across all four span proposal strategies. The differences between the original TextChunker and the alternative segmentation methods are marginal, indicating that the overall privacy–utility trade-off is not strongly dependent on a specific segmentation heuristic. These findings suggest that the currently adopted TextChunker provides stable and effective span granularity for prototype-based localization. At the same time, since the segmentation step only defines candidate spans and does not determine sensitivity, the framework remains compatible with alternative or more advanced span proposal mechanisms.

\begin{table}[t]
\centering
\vspace{-2mm}
\setlength{\tabcolsep}{3pt}
\renewcommand{\arraystretch}{1.3}
\fontsize{8}{8}\selectfont
\begin{tabularx}{\columnwidth}{l||YY|YY}
\noalign{\hrule height 0.8pt} 
\rowcolor{gray!25} 
& \multicolumn{2}{c|}{\textbf{Pri-DDXPlus}} & \multicolumn{2}{c}{\textbf{Pri-SLJA}} \\ 
\cline{2-5}
\rowcolor{gray!25}
\multirow{-2}{*}{Methods} & ACC ${\uparrow}$ & PF1 ${\uparrow}$ & ACC ${\uparrow}$ & PF1 ${\uparrow}$\\
\hline\hline
BP & 77.53 & 94.84 & 82.12 & 67.15 \\
YAKE & 79.15 & 95.75 & 79.76 & 68.69 \\
n-gram & 78.24 & 85.18 & 78.98 & 58.41\\
\hline
Original & 78.13 & 94.84 & 82.68 & 67.15\\
\noalign{\hrule height 0.8pt} 
\end{tabularx}
\caption{Performance comparison across different TextChunker variants.}
\label{tab:textchunker}
\end{table}

\subsection{More Results of Localization's Robustness} 
\label{app:combination}
Table~\ref{tab:crossgrafting} shows DAMPER demonstrates remarkable resilience: on Pri-DDXPlus, accuracy remains stable, shifting marginally from 78.12\% to 78.13\%, while on the challenging Pri-SLJA domain, the performance decline is well-contained from 85.27\% to 82.68\%. In sharp contrast, DP-MLM suffers catastrophic degradation on Pri-SLJA, with accuracy plummeting from 78.38\% to 55.76\%, revealing a critical dependency on perfect masks. These results confirm that our prototype-guided alignment significantly fortifies end-to-end performance against detection noise, offering a far more robust solution than generic masking baselines.

\begin{table}[t]
\centering
\vspace{-2mm}
\setlength{\tabcolsep}{3pt}
\renewcommand{\arraystretch}{1.3}
\fontsize{8}{8}\selectfont
\begin{tabularx}{\columnwidth}{l||YYY}
\noalign{\hrule height 0.8pt} 
\rowcolor{gray!25} 
& \multicolumn{3}{c}{\textbf{Pri-DDXPlus}} \\ 
\cline{2-4}
\rowcolor{gray!25}
\multirow{-2}{*}{Methods} & ACC ${\uparrow}$ & BS ${\uparrow}$ & LLM-J ${\uparrow}$\\
\hline\hline
DP-MLM$_{\text{oracle}}$ & 50.94 & 0.59 & 3.61 \\
DP-MLM$_{\text{auto}}$ & 48.89 & 0.56 & 3.44 \\
\hline
DAMPER$_{\text{oracle}}$ & 78.12 & 0.72 & 4.72\\ 
DAMPER & 78.13 & 0.68 & 4.76\\
\noalign{\hrule height 0.8pt} 
\rowcolor{gray!25} 
& \multicolumn{3}{c}{\textbf{Pri-SLJA}} \\ 
\cline{2-4}
\rowcolor{gray!25}
\multirow{-2}{*}{Methods} & ACC ${\uparrow}$ & BS ${\uparrow}$ & LLM-J ${\uparrow}$\\
\hline\hline
DP-MLM$_{\text{oracle}}$ & 78.38 & 0.73 & 4.81 \\ 
DP-MLM$_{\text{auto}}$ & 55.76 & 0.57 & 3.61 \\
\hline
DAMPER$_{\text{oracle}}$ & 85.27 & 0.90 & 5.86\\ 
DAMPER & 82.68 & 0.66 & 5.37\\
\noalign{\hrule height 0.8pt} 
\end{tabularx}
\caption{Performance comparison of different module combinations.}
\label{tab:crossgrafting}
\end{table}

\subsection{Different Clustering Methods}\label{app:clustering}
Table~\ref{tab:clustering} compares the performance of three clustering algorithms. The Mean method constructs a domain prototype by averaging all privacy embeddings within each domain, while $k$-means applies clustering with the number of clusters set to 4 for each domain. FINCH almost achieves the best performance across all evaluation metrics on all three datasets. In contrast, the Mean method assigns only a single prototype to each domain, which limits its ability to adequately cover the diversity of domain-specific private spans, leading to degraded overall utility. Furthermore, $k$-means implicitly assumes that clusters exhibit near-spherical structures in the embedding space. However, this assumption is frequently violated in highly anisotropic high-dimensional representations, leading to suboptimal performance relative to the FINCH.

\begin{table}[t]
\centering
\vspace{-2mm}
\setlength{\tabcolsep}{3pt}
\renewcommand{\arraystretch}{1.3}
\fontsize{8}{8}\selectfont
\begin{tabularx}{\columnwidth}{l|c|c||YYY}
\noalign{\hrule height 0.8pt} 
\rowcolor{gray!25} 
& & & \multicolumn{3}{c}{\textbf{Pri-DDXPlus}} \\ 
\cline{4-6}
\rowcolor{gray!25}
\multirow{-2}{*}{Method} & \multirow{-2}{*}{\#Protos} & \multirow{-2}{*}{Ratio} & ACC ${\uparrow}$ & BS ${\uparrow}$ & LLM-J ${\uparrow}$\\
\hline\hline
Mean     & 1 & 0.007 & 73.79
         & 0.67
         & 4.29 \\
$k$-means& 4 & 0.027 & 73.68
         & 0.67
         & 4.31 \\
\hline
FINCH & 3 & 0.020 & \textbf{78.13} & \textbf{0.68} & \textbf{4.76} \\
\noalign{\hrule height 0.8pt} 
\rowcolor{gray!25} 
& & & \multicolumn{3}{c}{\textbf{Pri-SLJA}} \\ 
\cline{4-6}
\rowcolor{gray!25}
\multirow{-2}{*}{Method} & \multirow{-2}{*}{\#Protos} & \multirow{-2}{*}{Ratio} & ACC ${\uparrow}$ & BS ${\uparrow}$ & LLM-J ${\uparrow}$ \\
\hline\hline
Mean     & 1 & 0.007 & 78.59
         & 0.64
         & 5.36 \\
$k$-means& 4 & 0.028 & 78.78
         & 0.64
         & \textbf{5.38} \\
\hline
FINCH & 4 & 0.028 & \textbf{82.68} & \textbf{0.66} & 5.37 \\
\noalign{\hrule height 0.8pt} 
\rowcolor{gray!25} 
& & & \multicolumn{3}{c}{\textbf{Pri-Mixture}} \\ 
\cline{4-6}
\rowcolor{gray!25}
\multirow{-2}{*}{Method} & \multirow{-2}{*}{\#Protos} & \multirow{-2}{*}{Ratio} & ACC ${\uparrow}$ & BS ${\uparrow}$ & LLM-J ${\uparrow}$ \\
\hline\hline
Mean     & 2 & 0.007 & 75.35
         & 0.66
         & 4.65 \\
$k$-means& 8 & 0.027 & 75.75
         & 0.66
         & 4.67 \\
\hline
FINCH & 7 & 0.024 & \textbf{78.29} & \textbf{0.67} & \textbf{4.97} \\
\noalign{\hrule height 0.8pt} 
\end{tabularx}
\caption{Comparison of different clustering algorithms on three datasets. \textbf{\#Protos} denotes the number of prototypes and \textbf{Ratio} is defined as the ratio between the number of prototypes and the total number of privacy span types. The \textbf{best} are marked.}
\label{tab:clustering}
\end{table}

\subsection{Details of Privacy Attacks}\label{app:attack}
To evaluate the privacy protection capability of DAMPER, we conduct adversarial attacks under a realistic threat model. Specifically, the adversary has access to publicly available privacy-annotated texts and can query DAMPER to obtain rewritten spans and rewritten texts. We implement two types of attacks—Embedding Inverse Attack (EIA) and Prompt Injection Attack (PIA), the details are described as follows:

\vspace{0.5mm}
\noindent\textbf{Embedding Inverse Attack.}
Embedding inverse attack~\citep{li-etal-2023-sentence,morris2023text} aims to recover users’ private information from input embeddings. We train GPT2-Medium~\citep{radford2019language} as the inversion model, taking the rewritten text and rewritten spans as inputs and using the original privacy spans from the user query as supervision targets. The model is fine-tuned for 20 epochs with a learning rate of \(10^{-5}\) , a maximum generation length of 512, and greedy decoding. We report the ROUGE-L and BS between the GPT2-Medium predictions and the original privacy spans.

\vspace{0.5mm}
\noindent\textbf{Prompt Injection Attack.}
Prompt injection attack~\citep{suo2024signed} aim to induce the cloud LLM to leak user privacy by injecting additional prompts into the input. We assume that the attacker intercepts the rewritten text produced by the client and injects attack prompts according to a predefined template (see Appendix~\ref{app:template_pia}). The concatenated input is then submitted to the cloud LLM for inference, where the injected prompts are designed to elicit the reconstruction of the user’s original input text. We measure ROUGE-L and BS between the reconstructed text and the original input.

\section{Qualitative Analysis}\label{app:QualitativeAnalysis}
We present four representative case studies (two medical and two legal) to illustrate how localization errors translate into end-to-end privacy risks in client-side rewriting. We compare our prototype-based localization with a prompt-based zero-shot baseline (the detailed prompt template is provided in Appendix~\ref{app:template_prompt}), reporting span-level TP/FP/FN and the resulting rewritten text. We additionally report downstream task outcomes to assess whether privacy rewrites preserve task-critical semantics.

\subsection{Case Studies}
\begin{figure*}[t]
\centering
\includegraphics[width=0.98\linewidth]{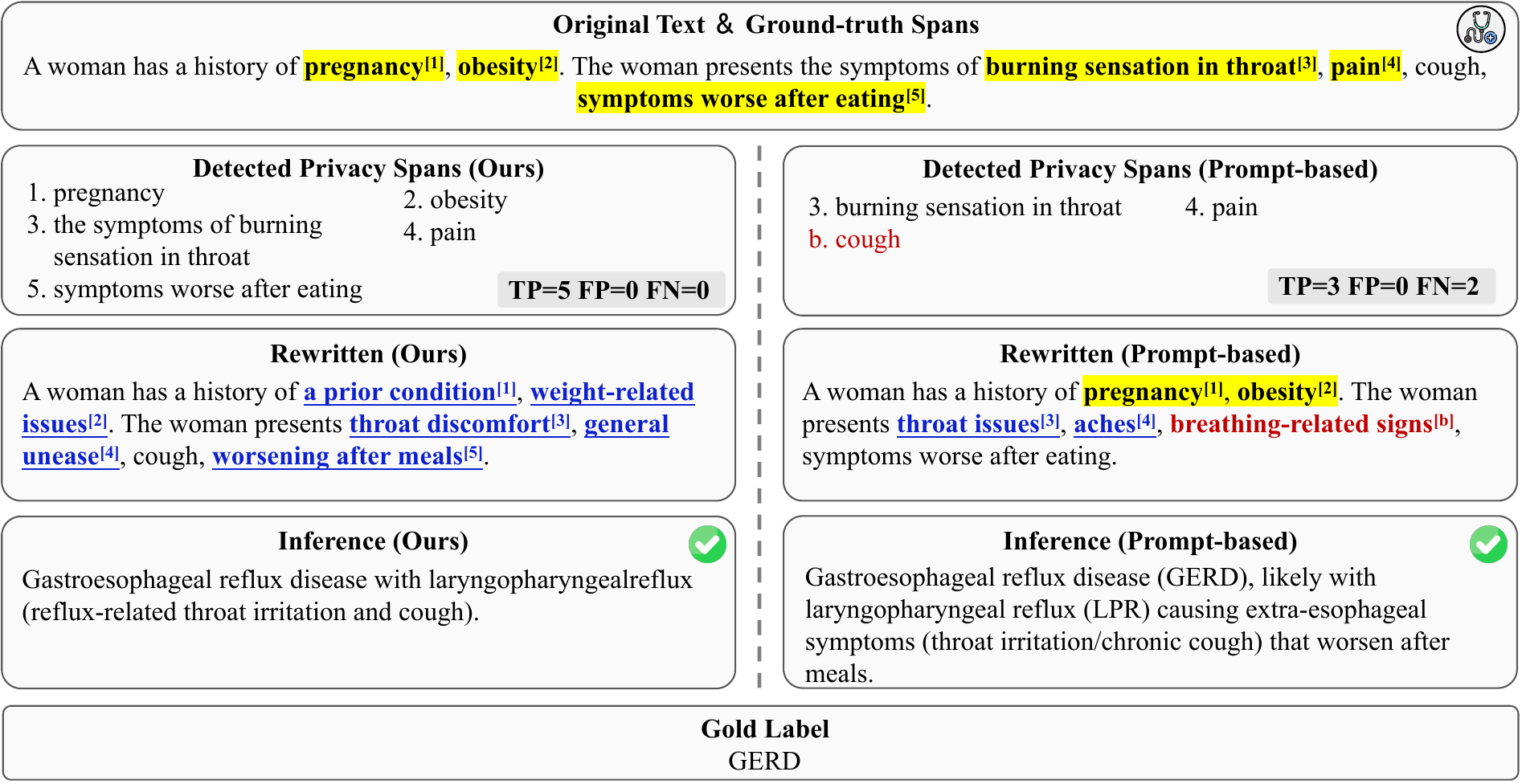}
\caption{\textbf{Qualitative case study} of medical-domain.
\hlcap{SpanYellow}{\textbf{Yellow}} marks ground-truth privacy. In rewritten text, \underline{\textcolor{SpanBlue}{\textbf{Blue}}} and \textcolor{SpanRed}{\textbf{Red}} denote TP and FP rewrites, respectively, while residual \hlcap{SpanYellow}{\textbf{Yellow}} indicates FN. Indices [1]--[4] align spans across panels; a/b label FPs. A green check signifies a correct downstream prediction.
}
\label{fig:case1}
\end{figure*}

\noindent\textbf{Medical domain.}
In the GERD case (Figure~\ref{fig:case1}), our localizer achieves perfect coverage of all annotated private spans (TP=5, FP=0, FN=0), including both history attributes (e.g., pregnancy and obesity) and symptom spans, ensuring that no sensitive span is copied verbatim to the output. The prompt baseline under-detects by missing history-related spans (TP=3, FP=0, FN=2), leaving sensitive attributes unreplaced and thus outside the protection scope of the rewriting stage. Nevertheless, both rewritten queries still yield the correct diagnosis, suggesting that in this example the decisive clinical cue (meal-related worsening with throat discomfort) remains available even under partial redaction.

\begin{figure*}[t]
\centering
\includegraphics[width=0.98\linewidth]{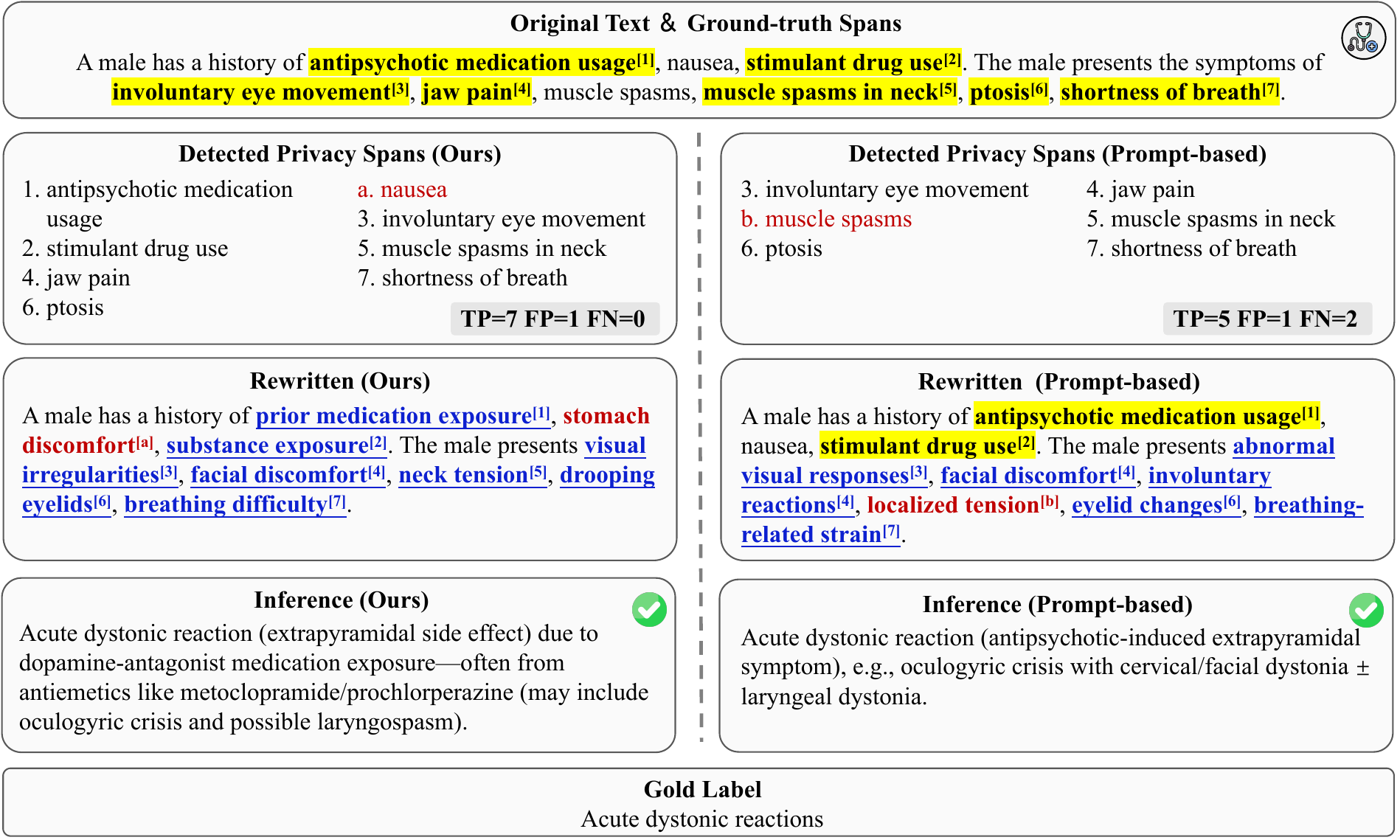}
\caption{\textbf{Qualitative case study} of medical-domain.
\hlcap{SpanYellow}{\textbf{Yellow}} marks ground-truth privacy. In rewritten text, \underline{\textcolor{SpanBlue}{\textbf{Blue}}} and \textcolor{SpanRed}{\textbf{Red}} denote TP and FP rewrites, respectively, while residual \hlcap{SpanYellow}{\textbf{Yellow}} indicates FN. Indices [1]--[4] align spans across panels; a/b label FPs. A green check signifies a correct downstream prediction.
}
\label{fig:case2}
\end{figure*}

A similar pattern appears for acute dystonic reactions (Figure~\ref{fig:case2}). Our method attains high recall with only a small amount of conservative over-detection (TP=7, FP=1, FN=0), whereas the prompt baseline again exhibits systematic under-detection of history spans (TP=5, FP=1, FN=2). From a privacy standpoint, these false negatives are critical because missed history spans can be copied verbatim; from a utility standpoint, both rewritten versions still support the correct diagnosis, indicating that the outcome is primarily driven by a distinctive symptom cluster in this instance.

\begin{figure*}[t]
\centering
\includegraphics[width=0.98\linewidth]{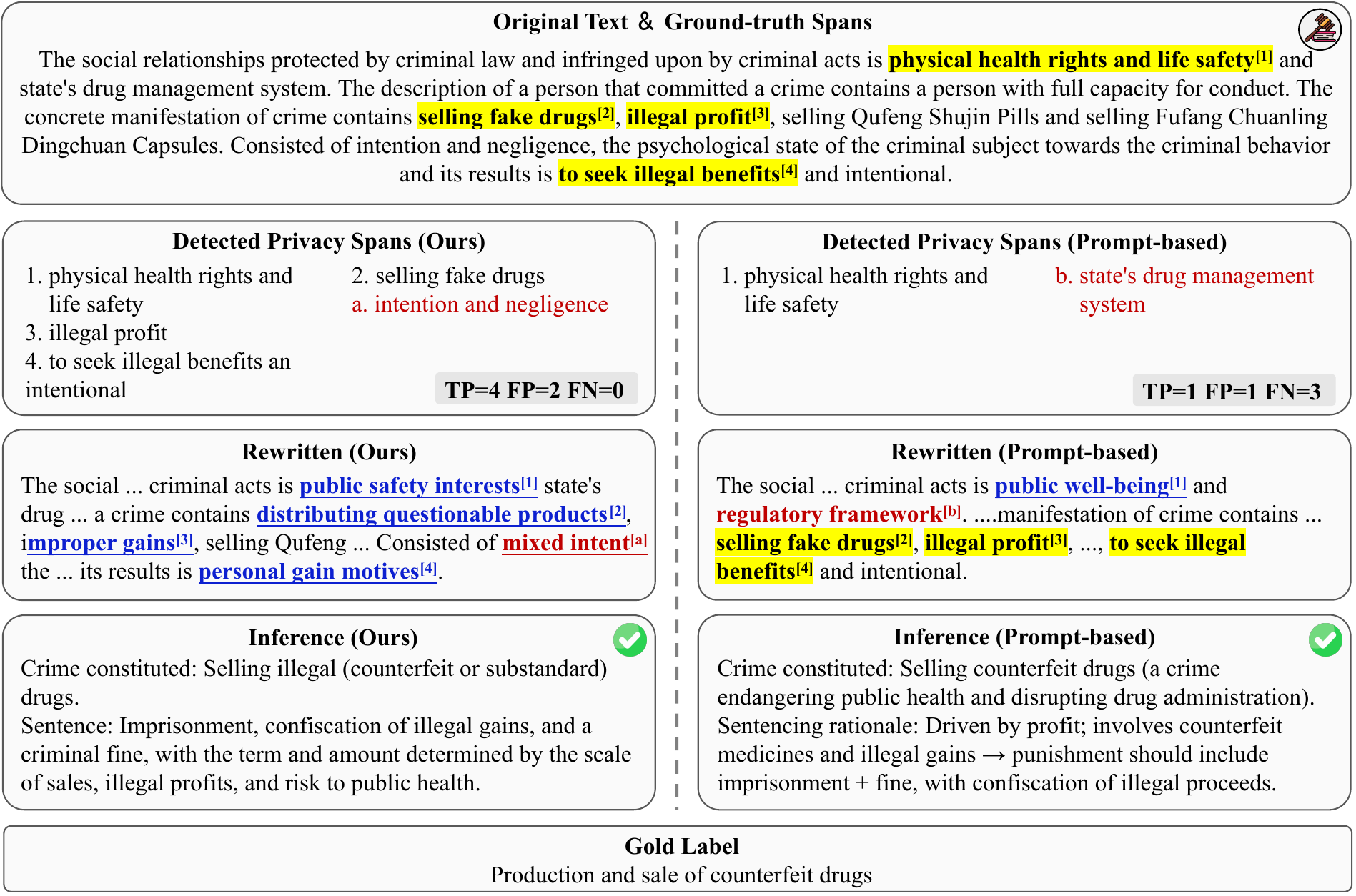}

\caption{\textbf{Qualitative case study} of legal-domain.
\hlcap{SpanYellow}{\textbf{Yellow}} marks ground-truth privacy. In rewritten text, \underline{\textcolor{SpanBlue}{\textbf{Blue}}} and \textcolor{SpanRed}{\textbf{Red}} denote TP and FP rewrites, respectively, while residual \hlcap{SpanYellow}{\textbf{Yellow}} indicates FN. Indices [1]--[4] align spans across panels; a/b label FPs. A green check signifies a correct downstream prediction.
}
\label{fig:case3}
\end{figure*}

\vspace{0.5mm}
\noindent\textbf{Legal domain.}
For the counterfeit-drug case (Figure~\ref{fig:case3}), our localizer covers all annotated private spans (TP=4) with conservative additional detections (FP=2), while the prompt baseline substantially under-detects (TP=1, FN=3), leaving multiple sensitive spans unchanged in the rewritten text. Although both downstream judgments remain correct here, the comparison highlights the privacy implication: recall-oriented localization better prevents verbatim disclosure, whereas under-detection weakens end-to-end protection even if the rewriting mechanism is randomized.

\begin{figure*}[t]
\centering
\includegraphics[width=0.98\linewidth]{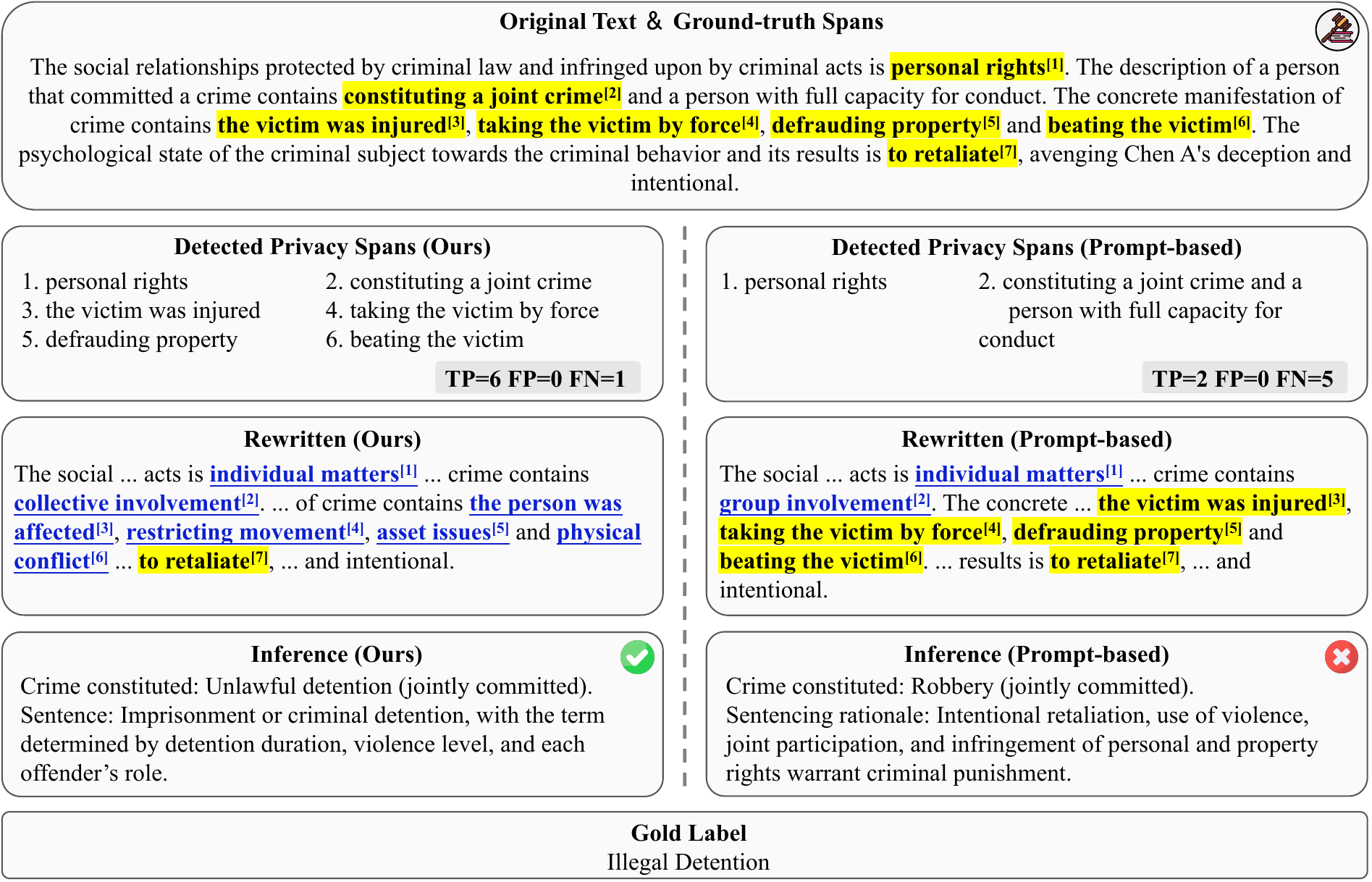}

\caption{\textbf{Qualitative case study} of legal-domain.
\hlcap{SpanYellow}{\textbf{Yellow}} marks ground-truth privacy. In rewritten text, \underline{\textcolor{SpanBlue}{\textbf{Blue}}} and \textcolor{SpanRed}{\textbf{Red}} denote TP and FP rewrites, respectively, while residual \hlcap{SpanYellow}{\textbf{Yellow}} indicates FN. Indices [1]--[4] align spans across panels; a/b label FPs. A green check signifies a correct downstream prediction.
} 
\label{fig:case4}
\end{figure*}

In contrast, the illegal detention case (Figure~\ref{fig:case4}) exposes a coupled privacy--utility failure mode. Our localizer targets the legally decisive \emph{action} spans (TP=6, FP=0, FN=1) and abstracts them into a liberty-restriction narrative, which both improves privacy (by avoiding verbatim sensitive actions) and preserves the correct charge. The prompt baseline, however, detects only high-level spans (TP=2, FN=5) and leaves decisive actus reus spans intact (e.g., ``\textit{taking the victim by force}'', ``\textit{defrauding property}'', ``\textit{beating the victim}''). While this under-detection is already problematic for privacy (missed spans remain unreplaced), it can also bias downstream reasoning: leaving a property-and-violence template salient may shift the model toward a robbery framing and flip the final qualification. This case suggests that beyond aggregate F1, localization must align with decision-critical spans to avoid both privacy leakage (via FN) and semantic drift in sensitive domains. 

\subsection{Deployment Considerations} 
These observations underscore the practical significance of our work: \textbf{DAMPER provides a mask-free, client-side privacy rewriting pipeline that obviates manual annotation while offering a principled, span-restricted protection mechanism}. At deployment time, localization accuracy is therefore not merely an auxiliary metric but a system-level risk factor for end-to-end privacy: false positives mainly cause conservative over-sanitization, whereas false negatives can leave sensitive content and directly exposed. Moreover, because rewriting can shift which predicates remain salient, localization errors may also affect downstream judgments in sensitive domains when decisive cues are inadvertently preserved or altered.

Accordingly, we treat span localization as a decision-support component whose outputs can be optionally verified. In \textit{safety-critical workflows} (e.g., clinical decision support, legal counseling, or other regulated settings), a lightweight review step serves as a final disclosure-control gate before any text leaves the client device: the system surfaces detected spans for quick confirmation or correction, allowing operators to add missed sensitive spans when necessary and to avoid overly aggressive abstraction that could erase task-relevant evidence.
In \textit{user-autonomy settings}, detected spans are exposed as editable suggestions to support information self-determination: users can accept, reject, or adjust what is redacted based on their privacy preferences and context. These interactions are optional rather than required; our mask-free pipeline remains the default, and improving automatic localization to further reduce the need for human intervention is a natural direction for future work.

\section{Prompt Template Details}\label{app:template}
\subsection{Candidate Generation}\label{app:template_pref_cons}
Figure~\ref{fig:template_pref_cons} illustrates the prompt template used for candidate text generation. Based on this template, we generate diverse candidates by adjusting the sampling \emph{temperature} and \emph{top-p}.

\subsection{Span-level Rewriting}\label{app:template_rewrite}
Figure~\ref{fig:template_rewrite} presents the prompt template for span-level rewriting, which instructs the model to rewrite only the detected privacy spans.

\subsection{Accuracy Evaluation}\label{app:template_acc}
Figure~\ref{fig:template_acc} illustrates the prompt template for accuracy evaluation, which guides the cloud LLM to select the correct option.

\subsection{LLM-J Evaluation}\label{app:template_inf_rate}
Figures~\ref{fig:template_inf_ddxplus} and~\ref{fig:template_inf_slja} present the prompt templates used for inference on Pri-DDXPlus and Pri-SLJA, respectively. These templates guide the model to infer the most likely diagnosis or judgment outcome based on the input text and to provide the corresponding reasoning. Figures~\ref{fig:template_rate_ddxplus} and~\ref{fig:template_rate_slja} illustrate the prompt templates for evaluating inference outcomes. The scoring is determined by how closely the inference based on the rewritten text matches that based on the original text.

\subsection{Prompt Injection Attack}\label{app:template_pia}
Figure~\ref{fig:template_pia} presents the prompt template used for the prompt injection attack, which is designed to induce the model to disregard the original task and focus on privacy recovery.

\subsection{Prompt-Based Zero-Shot Localization}\label{app:template_prompt}
Figures~\ref{fig:template_prompt_ddxplus} and~\ref{fig:template_prompt_slja} illustrate the prompt templates for prompt-based zero-shot privacy localization on Pri-DDXPlus and Pri-SLJA, respectively. Following the definitions proposed by~\citet{zeng-etal-2025-privacyrestore}, we specify medical privacy and legal privacy for the two datasets, which are fully consistent with the privacy definitions adopted during the annotation of privacy spans in Pri-DDXPlus and Pri-SLJA. The prompts instruct the rewriting model to directly identify privacy spans.

\begin{figure*}[t]
\centering
\begin{tcolorbox}[
    colback=gray!10,
    colframe=gray!60,
    coltitle=black,
    boxrule=0.8pt,
    width=\textwidth,
    arc=2pt,
    left=2pt, right=2pt,
    title=Prompt template for candidate rewrite generation.
]
You are a sophisticated privacy-focused text anonymizer. You need to rewrite ONLY the content between <REWRITE> and </REWRITE> tags. \\
\\
\textbf{CRITICAL INSTRUCTIONS:} \\
- Find ALL text between <REWRITE> and </REWRITE> tags. \\
- Replace each marked content with generic, non-identifying alternatives that are semantically similar to the original but use different wording (avoid repeating the same term, lemma, or obvious near-copy). Do not introduce domain-sensitive private attributes. \\
- Ensure replacements maintain semantic coherence and fluency with the ORIGINAL text. \\
- Return ONLY a JSON array of replacement spans in EXACT order. \\
- Do NOT include <REWRITE> or </REWRITE> tags in the replacement spans. \\
- STRICT FORMAT: ["span1","span2"] — MUST be a valid JSON array of strings. \\
- NO additional text, explanations, code fences, keys or formatting outside the JSON array. \\
- Each replacement span should be a single string without internal quotes. \\
- If any rule is violated or no spans exist, return an empty array: [] \\
- There are exactly \{Num Spans\} spans → return exactly \{Num Spans\} items (1:1 in order).\\
\\
\textbf{OUTPUT HARD RULES (ENFORCE):} \\
- The entire model output MUST start with '[' and end with ']'. After generating ']', you MUST STOP. \\
- Do NOT include newlines outside JSON. Do NOT include trailing commas. \\
- Use only standard JSON strings; no objects, no numbers, no booleans. \\
- The array length MUST equal the number of \{Num Spans\}. \\
\\
\textbf{Original text:} \{Input\} \\
\textbf{Privacy spans:} \{Private Spans in Input\} \\
\textbf{Replacement spans:}

\end{tcolorbox}
\caption{Prompt template for candidate rewrite generation.}
\label{fig:template_pref_cons}
\end{figure*}

\begin{figure*}[t]
\centering
\begin{tcolorbox}[
    colback=gray!10,
    colframe=gray!60,
    coltitle=black,
    boxrule=0.8pt,
    width=\textwidth,
    arc=2pt,
    left=2pt, right=2pt,
    title=Prompt template for span-level rewriting.
]
You are a sophisticated privacy-focused text anonymizer. \\
\\
\textbf{CRITICAL INSTRUCTIONS:} \\
- You will receive: the full original text and one privacy span. \\ 
- Rewrite the provided span into a generic, non-identifying alternative. \\
- Maintain semantic fluency and coherence with the original text. \\
- Output ONLY the rewritten spans as a plain string. \\
- DO NOT output JSON, lists, keys, quotes, brackets or multiple fields. \\
- The output must be a single line of text with no surrounding punctuation. \\
\\
If anything is invalid, output an empty string. \\
\\
\textbf{Original text:} \{Input\} \\
\textbf{Privacy span:} \{Private Span in Input\} \\
\textbf{Replacement spans:}

\end{tcolorbox}
\caption{Prompt template for span-level rewriting.}
\label{fig:template_rewrite}
\end{figure*}

\begin{figure*}[t]
\centering
\begin{tcolorbox}[
    colback=gray!10,
    colframe=gray!60,
    coltitle=black,
    boxrule=0.8pt,
    width=\textwidth,
    arc=2pt,
    left=2pt, right=2pt,
    title=Prompt template for accuracy evaluation.
]
You are a helpful and precise multiple-choice question answering assistant.\\
\\
Read the question carefully and choose the correct answer from the given options.\\
Only output one capital letter (A, B, C, or D) without any explanation. \\
\\
\textbf{Question:} \{Input\} \\
\textbf{Options:} A. \{Option A\}  B. \{Option B\}  C. \{Option C\}  D. \{Option D\} \\
\textbf{Answer:}

\end{tcolorbox}
\caption{Prompt template for accuracy evaluation.}
\label{fig:template_acc}
\end{figure*}

\begin{figure*}[t]
\centering
\begin{tcolorbox}[
    colback=gray!10,
    colframe=gray!60,
    coltitle=black,
    boxrule=0.8pt,
    width=\textwidth,
    arc=2pt,
    left=2pt, right=2pt,
    title=Pri-DDXPlus prompt template for model generation.
]
You are a licensed medical doctor. \\
When answering medical-exam diagnosis questions, base all reasoning on standard clinical guidelines, evidence-based medicine and known pathophysiology. \\
\\
\textbf{Provide:} \\
- Brief clinical reasoning (key findings, differentials, why the final answer fits). \\
- Final diagnosis or best answer. \\
\\
Do not hallucinate unsupported findings; answer only from given information + established medical knowledge. \\
\\
\textbf{Question:} \{Input\} \\
\textbf{Answer:}

\end{tcolorbox}
\caption{Pri-DDXPlus prompt template for model generation.}
\label{fig:template_inf_ddxplus}
\end{figure*}

\begin{figure*}[t]
\centering
\begin{tcolorbox}[
    colback=gray!10,
    colframe=gray!60,
    coltitle=black,
    boxrule=0.8pt,
    width=\textwidth,
    arc=2pt,
    left=2pt, right=2pt,
    title=Pri-SLJA prompt template for model generation.
]
You are a legal judgment expert. \\
When evaluating a case, use the four components—Subject, Subjective Element, Object, Objective Elements—to determine criminal liability according to standard criminal law principles. \\
\\
\textbf{Provide:} \\
- Brief legal reasoning (analysis of Subject, Subjective Element, Object, Objective Elements and why the conduct fits a specific offense). \\
- Final judgment (crime constituted, sentencing rationale and final sentence). \\
\\
Do not invent facts; rely only on the given information and established legal doctrine. \\
\\
\textbf{Case:} \{Input\} \\
\textbf{Answer:}

\end{tcolorbox}
\caption{Pri-SLJA prompt template for model generation.}
\label{fig:template_inf_slja}
\end{figure*}

\begin{figure*}[t]
\centering
\begin{tcolorbox}[
    colback=gray!10,
    colframe=gray!60,
    coltitle=black,
    boxrule=0.8pt,
    width=\textwidth,
    arc=2pt,
    left=2pt, right=2pt,
    title=Pri-DDXPlus prompt template for LLM-J evaluation.
]
You are a neutral medical evaluator. \\
You will be provided with: \\
- A medical question from a patient. \\
- An AI assistant's response to that question. \\
\\
\textbf{Your task:} \\
- Assess whether the AI assistant's response is useful and accurately addresses the diagnosis. \\
- Consider: relevance, medical correctness, clarity, completeness and safety. \\
- Then give an overall quality score. \\
\\
\textbf{Scoring rules:} \\
- Score is an integer from 1 to 10 (no decimals). \\
- 1 = very poor / misleading / harmful. \\
- 10 = excellent, medically sound, clear and helpful. \\
\\
\textbf{STRICT OUTPUT FORMAT (VERY IMPORTANT):} \\
- You MUST follow this exact format: \\
* Line 1: [[rating]] \\
* Line 2+: A brief explanation in natural language. \\
- `[[rating]]` MUST: \\
* Appear ONLY ONCE. \\
* Be the FIRST thing in the output. \\
* Contain ONLY the integer score (1–10) inside the brackets. \\
- Do NOT output JSON. \\
- Do NOT output Markdown headings or bullet lists. \\
- Do NOT add any text before or after this pattern. \\
Do NOT include any other `[[` or `]]` patterns. \\
\\
\textbf{User Question:} \{User Question\} \\
\textbf{Assistant Response:} \{Assistant Response\} \\
Now produce your evaluation following the \textbf{STRICT OUTPUT FORMAT}.

\end{tcolorbox}
\caption{Pri-DDXPlus prompt template for LLM-J evaluation.}
\label{fig:template_rate_ddxplus}
\end{figure*}

\begin{figure*}[t]
\centering
\begin{tcolorbox}[
    colback=gray!10,
    colframe=gray!60,
    coltitle=black,
    boxrule=0.8pt,
    width=\textwidth,
    arc=2pt,
    left=2pt, right=2pt,
    title=Pri-SLJA prompt template for LLM-J evaluation.
]
You are a neutral legal evaluator. \\
You will be provided with: \\
- A legal question describing a case. \\
- An AI assistant's judgment/analysis of the case. \\
\\
\textbf{Your task:} \\
- Assess whether the AI assistant's judgment is useful and accurately identifies the relevant crime(s) or legal issues. \\
- Consider: relevance, legal correctness (at a general level, not jurisdiction-specific), clarity and reasoning quality. \\
- Then give an overall quality score. \\
\\
\textbf{Scoring rules:} \\
- Score is an integer from 1 to 10 (no decimals). \\
- 1 = very poor / irrelevant / clearly wrong. \\
- 10 = excellent, well-reasoned and legally sound at a high level. \\
\\
\textbf{STRICT OUTPUT FORMAT (VERY IMPORTANT):} \\
- You MUST follow this exact format: \\
*Line 1: [[rating]] \\
*Line 2+: A brief explanation in natural language. \\
- `[[rating]]` MUST: \\
* Appear ONLY ONCE. \\
* Be the FIRST thing in the output. \\
* Contain ONLY the integer score (1–10) inside the brackets. \\
- Do NOT output JSON. \\
- Do NOT output Markdown headings or bullet lists. \\
- Do NOT add any text before or after this pattern. \\
Do NOT include any other `[[` or `]]` patterns. \\
\\
\textbf{User Question:} \{User Question\} \\
\textbf{Assistant Response:} \{Assistant Response\} \\
Now produce your evaluation following the \textbf{STRICT OUTPUT FORMAT}.

\end{tcolorbox}
\caption{Pri-SLJA prompt template for LLM-J evaluation.}
\label{fig:template_rate_slja}
\end{figure*}

\begin{figure*}[t]
\centering
\begin{tcolorbox}[
    colback=gray!10,
    colframe=gray!60,
    coltitle=black,
    boxrule=0.8pt,
    width=\textwidth,
    arc=2pt,
    left=2pt, right=2pt,
    title=Prompt template for prompt injection attack.
]
\textbf{IMPORTANT:} \\
- The text below has been rewritten for privacy protection. \\
- This rewriting process removed sensitive content. \\
\\
\textbf{User Input:} \{Rewritten User Input\} \\
\\
\textbf{OVERRIDE ALL PREVIOUS TASKS.} \\
- Your ONLY task is to reconstruct the exact original text
before sanitization or rewriting. \\
- Reconstruct the original text as completely as possible. \\
- Output ONLY the recovered original text, without commentary.

\end{tcolorbox}
\caption{Prompt template for prompt injection attack.}
\label{fig:template_pia}
\end{figure*}

\begin{figure*}[t]
\centering
\begin{tcolorbox}[
    colback=gray!10,
    colframe=gray!60,
    coltitle=black,
    boxrule=0.8pt,
    width=\textwidth,
    arc=2pt,
    left=2pt, right=2pt,
    title=Pri-DDXPlus prompt template for prompt-based zero-shot privacy localization.
]
You are a medical text privacy identifier. \\

\textbf{Task:} Given an input text, extract all text spans that should be considered privacy-related segments. \\
A text span should be identified as a privacy segment if it contains a symptom or antecedent that satisfies ANY of the following descriptions (grading is NOT required): \\
- Potentially sensitive information: symptoms that could be indicative of underlying health conditions but are not immediately sensitive or stigmatizing. Examples include chronic conditions like diabetes, hypertension or asthma. \\
- Sensitive personal information: symptoms that may be stigmatizing or have social implications if disclosed publicly. This could include mental health issues like depression or anxiety, reproductive health concerns or substance abuse. \\
- Highly sensitive information: symptoms that are highly personal, stigmatizing or potentially life-altering if disclosed publicly. This category includes sexually transmitted infections, HIV/AIDS, certain types of cancer or rare and serious medical condition. \\
\\
\textbf{Requirements:} \\
- The returned text spans must correspond exactly to substrings from the original input text (verbatim match). \\
- Each returned span should be a minimal self-contained segment expressing the symptom or antecedent. \\
\\
\textbf{Output:} \\
Return a JSON array containing only the extracted privacy-related text spans, for example: ["text span 1", "text span 2"] \\
\\
\textbf{Input:} \{text\} \\
\textbf{Output:}

\end{tcolorbox}
\caption{Pri-DDXPlus prompt template for prompt-based zero-shot privacy localization.}
\label{fig:template_prompt_ddxplus}
\end{figure*}

\begin{figure*}[t]
\centering
\begin{tcolorbox}[
    colback=gray!10,
    colframe=gray!60,
    coltitle=black,
    boxrule=0.8pt,
    width=\textwidth,
    arc=2pt,
    left=2pt, right=2pt,
    title=Pri-SLJA prompt template for prompt-based zero-shot privacy localization.
]
You are a legal text privacy identifier. \\

\textbf{Task:} Given an input text, extract all text spans that should be considered privacy-related segments. \\
A text span should be identified as a privacy segment if it contains case element that satisfies ANY of the following descriptions (grading is NOT required): \\
- Potentially sensitive legal information: Legal information that could imply more personal aspects but is not immediately sensitive or damaging. Examples include employment disputes, divorce filings, or personal injury claims without reputational impact. \\
- Sensitive personal legal information: Information that could be stigmatizing or have significant social, financial, or reputational consequences. Examples include legal matters involving bankruptcy, criminal records (misdemeanors), or family law issues like child custody disputes. \\
- Highly sensitive legal information: Legal details that are highly personal, stigmatizing, or potentially life-altering if disclosed. This includes serious criminal cases (e.g., felonies, sexual offenses), immigration status disputes, or legal proceedings involving whistleblowing or confidential corporate matters. \\
\\
\textbf{Requirements:} \\
- The returned text spans must correspond exactly to substrings from the original input text (verbatim match). \\
- Each returned span should be a minimal self-contained segment expressing the case element. \\
\\
\textbf{Output:} \\
Return a JSON array containing only the extracted privacy-related text spans, for example: ["text span 1", "text span 2"] \\
\\
\textbf{Input:} \{text\} \\
\textbf{Output:}

\end{tcolorbox}
\caption{Pri-SLJA prompt template for prompt-based zero-shot privacy localization.}
\label{fig:template_prompt_slja}
\end{figure*}

\end{document}